\documentclass[11pt,letter,subeqn,fleqn]{article}
\usepackage{amsmath, amssymb,amsthm,cite}
\usepackage[dvips]{graphics,color}
\usepackage{subfig}
\usepackage{graphicx}
\usepackage{tabulary}
\usepackage{float}
\usepackage{xcolor}
\usepackage[utf8]{inputenc}
\usepackage{hyperref}

\allowdisplaybreaks

\title{Invariant Conservation Law-Preserving Discretizations of Linear and Nonlinear Wave Equations}

\author{ 
~A. F. Cheviakov\footnotemark[1],
~V. A. Dorodnitsyn\footnotemark[2],
~E. I. Kaptsov\footnotemark[3]\\
\small $^{\rm a}$\emph{Department of Mathematics and Statistics, University of Saskatchewan, Saskatoon, Canada}\vspace{0.2cm}\\
\small $^{\rm b,c}$\emph{Keldysh Institute of Applied Mathematics of Russian Academy of Science,}\\
\emph{Miusskaya Pl. 4, Moscow, 125047}\vspace{0.2cm}\\
}

\def\beq{\begin{equation}}
\def\eeq{\end{equation}}
\def\barr{\begin{array}{ll}}
\def\barrcl{\begin{array}{rcl}}
\def\earr{\end{array}}
\def\eps{\varepsilon}

\def\sg#1{{\rm #1}}
\def\const{\hbox{\rm const}}

{\theoremstyle{definition} 

\newtheorem{remark}{Remark}
}

\newcounter{tabnum}

\def\dh#1{\mathop {#1}\limits_{-h}}
\def\dhp#1{\mathop {#1}\limits_{+h}}
\def\dtau#1{\mathop {#1}\limits_{-\tau}}
\def\dtaup#1{\mathop {#1}\limits_{+\tau}}

\def\DHP{\dhp{D}}
\def\DHM{\dh{D}}
\def\DTP{\dtaup{D}}
\def\DTM{\dtau{D}}
\def\SHP{\dhp{S}}
\def\SHM{\dh{S}}
\def\STP{\dtaup{S}}
\def\STM{\dtau{S}}

\newcommand{\WNL}{\mathcal{W}_{\text{NL}}}

\begin{document}


\footnotetext[1]{Corresponding author. Alternative English spelling: Alexey Shevyakov. Electronic mail: a.shevyakov@usask.ca}
\footnotetext[2]{Electronic mail: dorod@spp.Keldysh.ru}
\footnotetext[3]{Electronic mail: evgkaptsov@gmail.com}

\maketitle \numberwithin{equation}{section}
\maketitle \numberwithin{remark}{section}
\numberwithin{lemma}{section}
\numberwithin{proposition}{section}

\begin{abstract}

Symmetry- and conservation law-preserving finite difference discretizations are obtained for linear and nonlinear one-dimensional wave equations on five- and nine-point stencils, using the theory of Lie point symmetries of difference equations, and the discrete direct multiplier method of conservation law construction. In particular, for the linear wave equation, an explicit five-point scheme is presented that preserves the discrete analogs of its basic geometric point symmetries, and six of the corresponding conservation laws. For a class of nonlinear wave equations arising in hyperelasticity, a nine-point implicit scheme is constructed, preserving four point symmetries and three local conservation laws. Other discretization of the nonlinear wave equations preserving different subsets of conservation laws are discussed.

\end{abstract}

\section{Introduction}\label{sec:Intro}

Symmetries are a fundamental intrinsic feature of
differential equations of mathematical physic. Lie groups of local symmetries yield a number
of useful geometrical properties of differential equations (see
\cite{ovsiannikov2014group,OlverBk,ibragimov2001transformation,BlumanAncoDEs2002, BCAbook}).
For ordinary differential equations (ODE), the invariance with respect to Lie
group of transformations gives a possibility of a reduction of order (possibly to complete integration if there is a sufficient number of symmetries),
derivation of families of new solutions from a given one, first integrals,
etc. For partial differential equations (PDE), their symmetries and
conservation laws are the related parts of their coordinate-invariant
structure, containing essential analytical
information. In addition to a direct physical interpretation,
symmetries and conservation laws are used for establishing S- and
C-integrability, mappings between equations, construction of exact
invariant solutions; they are used in proofs of existence,
uniqueness and stability of solutions, derivation of exact
closed-form solutions, and other purposes (see, e.g.,
Refs.~\cite{OlverBk, BCAbook} and references therein). For models
arising from a variational principle, the second Noether's theorem
relates local variational symmetries and local conservation laws;
for non-variational models, other relationships exist \cite{OlverBk,
BCAbook, anco2017generalization}. Conservation laws are key elements
in some finite-volume and finite-element numerical methods (e.g.,
\cite{leveque1992numerical, kallendorf2012conservation}). In
addition to local symmetries and conservation laws, some models admit nonlocal symmetries and
conservation laws, which may be computed using various approaches
\cite{krasilshchik1984nonlocal, bocharov1999symmetries, BCI2,
BCAbook, cheviakov2017recursion}.

Over the last 25 years, considerable progress has been made  in the
applications of the theory of Lie groups of transformations and related methods to difference equations. (See, e.g., \cite{[Pavel], [Pavel2], VDGroups, [Dor_4], [Dor_5], [Dheat],
[D_K_W_1], [Hydon-Mansfield], [LW-4], [LW-5], [LW-6], [LW-11],
[LW-12], olver2001geometric, [Quispel-1]}, and reviews in~\cite{[D-book], [LW-2], [LW-3]}.)

If no invariant Lagrangian or Hamiltonian exists the alternative
methods of constructing lower order first integrals have been
proposed in~\cite{AB97, BlumanCheviakovAncoPDEs} and \cite{ibragimov2007new,[ibr11a], [Ibr]}. They make use of the so called adjoint equation
solutions of which one uses to construct the required first
integrals. The crucial point of this ``adjoint equation method" is the Lagrange operator
identity which connects symmetry of a given equation with
conservation laws and solutions of adjoint equation. The difference analog of the
adjoint equation method based on newly established difference analog
of the Lagrange identity was applied to ordinary difference
equations in~\cite{[DorKapKozWint]}. An alternative ``direct method" (or ``multiplier method") of local conservation law
computation \cite{alonso1979noether,
vinogradov1984local, OlverBk, Anco1997,Anco2002part1,BCAbook} is based on the characteristic form of the divergence
expressions, and employs Euler differential operators (variational
derivatives). The direct method has been used to compute
conservation laws for multiple models (see, e.g.,
Refs.~\cite{BCAbook, BC1, BCI2, bluman2007nonlocally,
bluman2008nonlocally, kelbin2013new, kallendorf2012conservation,
cheviakov2014generalized, kunzinger2008potential,
anco2010conservation, kelbin2013new}); it was implemented in
symbolic software \cite{GeM2007,C_flux,cheviakov2010symbolic,
cheviakov2017symbolic}. The use of the direct method has been shown to be often
computationally advantageous even for variational models
\cite{BCApap, BCAbook, cheviakov2015comparison}.

If a given ODE or PDE system is solved numerically using a
finite-difference method, the choice of a mesh and the
finite-difference approximation itself depend on the desired order
of approximation, specific features of the model, such as domain
geometry, and other considerations, such as simplicity of
implementation, computation speed, absolute stability, etc. In many
cases, there are multiple discretizations on the same stencil, with
the same order of approximation; the choice of a particular numerical scheme
may then be based on analytical properties of the difference equations.

Symmetries, first integrals and conservation laws for differential equations have discrete analogs for difference equations (see, for example, Refs.~\cite{VDBook, VDGroups, dorodnitsyn2013discretization}, and references therein). Other related properties, such as discrete variational and Hamiltonian formulations and Noether's theorem, have also been established \cite{veselov1991integrable, ablowitz1996numerical, marsden1998multisymplectic, dorodnitsyn2001noether, dorodnitsyn2010invariance}. Much work has been recently done in terms of development of numerical algorithms that respect analytical properties of the model, in general, as well as for specific models. Examples of symmetry-preserving numerical schemes, conservation law-preserving, symplectic and multisymplectic discretizations are known for many models (such as Refs. \cite{ablowitz1996numerical, marsden1998multisymplectic, olver2001geometric, dorodnitsyn2003heat, chen2003multisymplectic, de2004geometric, valiquette2005discretization, olver2009lectures, rebelo2013symmetry, bihlo2017symmetry}, to name a few). Nonlocal difference conservation laws for ODEs were derived in Ref.~\cite{dorodnitsyn2001noether}.
Conservation law-preserving schemes have been analyzed  for parabolic PDEs \cite{dorodnitsyn2003heat}, one-dimensional gas dynamics equations~\cite{dorodnitsyn2018one},
shallow water equations~\cite{dorodnitsyn2019shallow,bk:Bihlo_numeric[2012]}, and other models. Conversely, as it has been pointed out, for example, in \cite{samarskii1980difference}, if a numerical scheme does not preserve a discrete analog of a physical conservation law, then such a discretization will include artificial sources of the conserved quantity that do not have a physical meaning. The imbalance generated by such fictitious sources accumulates over time, and might not be removed even by the spatial mesh refinement. Such artificial source terms often involve derivatives of scalar fields, which results in fast error growth on solutions quickly fluctuating in space and/or time even on a fine mesh.

In Ref.~\cite{wan2016multiplier} it was shown that if a conservation law for a PDE system in the characteristic form is discretized, it provides a consistent discretization for one of the PDEs of the given model, having the same accuracy; moreover, an multiplier method-based algorithm to yield conservation-law preserving discretizations of one DE with a single conservation law, and a system of $m$ PDEs with $s\leq m$ conservation laws, was provided.

\medskip
In the current contribution, we focus on the application of the discrete analog of the direct multiplier method to systematically compute difference-type conservation laws for numerical discretizations of hyperbolic PDE models, and derive finite-difference numerical schemes that preserve such conservation laws. Similarly to its continuous analog, the discrete direct method uses the mesh version of Euler differential operators to find multiplies; the latter, on a given stencil, convert a system difference equations into a discrete divergence expression. We use this algorithm to derive invariant discretizations for that admit discrete analogs of continuous conservation laws of the models. In particular, unlike the case in Ref.~\cite{wan2016multiplier}, we are interested in constructing discretizations that would exactly preserve discrete analogs of multiple conservation laws holding for a single PDE.

As the main example, we consider a class of nonlinear one-dimensional wave equations arising in models of finite (non-small) shear displacements in an anisotropic hyperelastic solid containing a family of elastic fibers \cite{cheviakov2016one}. In particular, these models describe elongated soft biological tissues such as muscles. Such models use Mooney-Rivlin-type strain energy density with an additional quadratic fiber-dependent term; the fiber family is assumed to have a constant material direction, making a constant angle $\gamma$, $|\gamma|\leq \pi/2$, with the direction $x$ of wave propagation. The assumption of displacements $G(x,t)$ being  transverse to material direction $x$ of wave propagation yields an incompressible model given by a single PDE
\beq\label{eq:anz:1fib:hyperel:gengam:PDE}
G_{tt} = \left( \alpha
    + {\beta\cos^{2}\gamma  \left( 3\cos^{2}\gamma  \left(G_x\right)^{2} + 6\sin\gamma \cos\gamma  \,G_x + 2\sin^2\gamma  \right) }\right)  G_{xx}.
\eeq
where $\alpha, \beta>0$ are constant material and fiber strength parameters. When $\beta\to 0$, the model becomes fiber-independent; when $\gamma=\pi/2$, the fibers are perpendicular to the wave propagation direction, and play no role \cite{cheviakov2015fully}. In both of these cases, the nonlinear model reduces to the linear wave equation
\beq\label{eq:lin:Wave:intro}
G_{tt} = \alpha G_{xx},
\eeq
which through a rescaling can be written as $G_{tt} = G_{xx}$. In another case when the fibers are parallel to the wave propagation direction ($\gamma=0$), the PDE \eqref{eq:anz:1fib:hyperel:gengam:PDE} takes the simple form
\beq\label{eq:anz:1fib:hyperel:gam0:PDE}
G_{tt} = (\alpha + 3\beta G_x^2)\,G_{xx},
\eeq
Moreover, it was shown \cite{cheviakov2016one} that equivalence transformations that can be used to map the general PDE \eqref{eq:anz:1fib:hyperel:gengam:PDE} ($\gamma\ne 0$) into the form \eqref{eq:anz:1fib:hyperel:gam0:PDE}. While the general closed-form solution of the PDE \eqref{eq:anz:1fib:hyperel:gam0:PDE} is unknown, its conservation laws and numerical solutions were studied in Ref.~\cite{cheviakov2016one}. We note that the general PDE \eqref{eq:anz:1fib:hyperel:gengam:PDE} as well as its specific cases \eqref{eq:lin:Wave:intro}, \eqref{eq:anz:1fib:hyperel:gam0:PDE} admit variational formulations.


\medskip
The current contribution is organized as follows.

In Section \ref{sec:2}, we review the notions of symmetry invariance, local conservation laws, and their direct construction using multipliers, for both partial differential equations and their finite-difference discretizations.

Section \ref{sec:linW} is devoted to the first, simplest example of a hyperbolic PDE: a linear wave equation \eqref{eq:lin:Wave:intro} in 1+1 dimensions. This equation is invariant with respect to an infinite set of Lie point symmetries, including several basic geometric symmetries: three shifts, a Galilei transformation, a stretch, a Lorentzian boost, and two scalings.  Seven of these symmetries are variational; the corresponding conservation laws are easily computed using Noether's theorem or the direct method. We next show that the five-point cross-type symmetric second-order finite difference discretization of the linear wave equation on a uniform mesh is both a symmetry- and a conservation law-preserving discretization. In particular, it admits discrete analogs of all of the above geometric symmetries with the exception of the Lorentz transformation (which breaks the mesh orthogonality). Moreover, the symmetric cross-stencil discretization admits a discrete variational formulation, and has discrete analogs of six of the seven conservation laws. We show that there is no scheme on the same five-point stencil that would preserve the remaining conservation law.

In Section \ref{sec:nonlinW}, the nonlinear wave equation \eqref{eq:anz:1fib:hyperel:gam0:PDE} is considered. After using equivalence transformations to remove arbitrary constants, we present the five point symmetries of the model (three translations, Galilei transformation, and a scaling), the Lagrangian density, and four local conservation laws (corresponding to all symmetries but the scaling). For the simplest, explicit  five-point finite-difference discretization of the model PDE on the uniform orthogonal mesh, we show that analogs of only two of the four conservation laws hold. The next step is to come up with a ``better" scheme on a more general nine-point stencil. We show that there exists an explicit discretization which is invariant with respect to admitting analogs of three conservation laws, and there does not exist a polynomial scheme preserving all four conservation laws of the nonlinear wave equation. (We show, however, that other numerical schemes preserving specifically the ``missing" conservation law can be constructed.)


\medskip
Throughout the paper, we use subscripts
\[
u_t\equiv \frac{\partial}{\partial t} u(t,x),\quad u_{xx}\equiv \frac{\partial^2}{\partial x^2} u(t,x),
\]
in terms of independent variables to denote the corresponding partial derivatives. Without ambiguity, subscripts involving integer indices are used for indexing discrete sets: $x_m$, $t_n$, $U_{m}^{n}$, etc. \verb|GeM| symbolic software package for \verb|Maple| \cite{GeM2007,C_flux,cheviakov2010symbolic,
cheviakov2017symbolic} was used for symmetry and conservation law computations.


\section{Invariant finite-difference schemes for partial differential equations and difference conservation laws}\label{sec:2}

\subsection{ PDEs and their finite-difference approximations}\label{sec:2:not}

Let
\beq\label{eq:PDEgen}
\Delta:~~R^\sigma[u]=R^\sigma(x,u,\partial u,\ldots,\partial^k u)=0,
\quad
\sigma=1,\ldots ,K
\eeq
denote a system of $K$ PDEs of order up to $k$, with $p$ independent variables $x=(x^1,\ldots ,x^p)$ and $q$ dependent variables $u(x)=(u^1(x)$, $\ldots$, $u^q(x))$. In \eqref{eq:PDEgen}, $\partial^s u$ is the set of all partial derivatives of $u$ of order $s$. The solution manifold of the PDE system \eqref{eq:PDEgen} is a set in the jet space  $\mathcal{J}^k(x|u)$ of variables $(x,u, \partial u, \ldots,\partial^k u)$. The bracket notation $R^\sigma[u]$ in \eqref{eq:PDEgen} and below is used to denote differential functions depending on the jet space variables to some fixed differential order.

In the current paper, we will work with single second-order PDEs of the form
\beq\label{eq:PDE2}
H[u] \equiv H(t,x,u,u_t,u_x, u_{tt},u_{tx},u_{xx})=0,
\eeq
involving  a single dependent variable $u(t,x)$. For a finite-difference numerical scheme approximating PDEs \eqref{eq:PDE2}, the mesh points are denoted by
\begin{equation}\label{eq:mesh}
  (t_n, x_m),\quad m=0,\ldots, M, \quad n=0,\ldots, N.
\end{equation}
For a general stationary mesh, the variable mesh steps are given by
\begin{equation}\label{eq:mesh:stationary}
\begin{array}{ll}
  h_{m} = x_{m+1} - x_{m}, &m=0,\ldots, M-1,\\[1ex]
  \tau_{n} =t_{n+ 1} - t_{n}, & n=0,\ldots, N-1.
\earr
\end{equation}
The dependent variable $u(t,x)$ at the mesh node $(t_n, x_m)$ is approximated by the discrete mesh quantity $U_{m}^{n}$:
\[
u(t_n,x_m)\simeq U_{m}^{n},\quad m=0,\ldots, M-1,\quad n=0,\ldots, N-1.
\]

Let $(\mathbf {t}, \mathbf {x}, \mathbf{U})$ denote the stencil of a finite-difference numerical method, a subset of the mesh \eqref{eq:mesh}. The specific type of the stencil is not pre-determined; it is chosen separately for each numerical scheme. On the stencil, the spatial coordinate $x$ and the corresponding index $m$ change along the horizontal axis, while the time coordinate $t$ and the corresponding index $n$ change along the vertical axis; here $m,n\in \mathbb{Z}$.
For a PDE \eqref{eq:PDE2}, a general finite-difference scheme has the form
\begin{equation}  \label{schemeGeneral}
\Delta_h:~~ \begin{cases}
  F(\mathbf{t}, \mathbf{x}, \mathbf{U}) = 0, \\
  \Omega^1(\mathbf{t}, \mathbf{x}, \mathbf{U}) = 0,\\
  \Omega^2(\mathbf{t}, \mathbf{x}, \mathbf{U}) = 0,
\end{cases}
\end{equation}
where
\beq\label{eq:diff:F}
F(\mathbf{t}, \mathbf{x}, \mathbf{U})=0
\eeq
is a \emph{partial difference equation} (P$\Delta$E) approximating the PDE  \eqref{eq:PDE2}, and the relations  ${\Omega}^1=0$, ${\Omega}^2=0$ are the \emph{mesh equations}. In the limit $(h \to 0, \tau \to 0)$ the mesh equations reduce to the identities $0=0$ (see \cite{WinternValiq2005, [D-book]} for details).

\begin{figure}[htbp]
\begin{center}
\includegraphics[width = 7cm,clip]{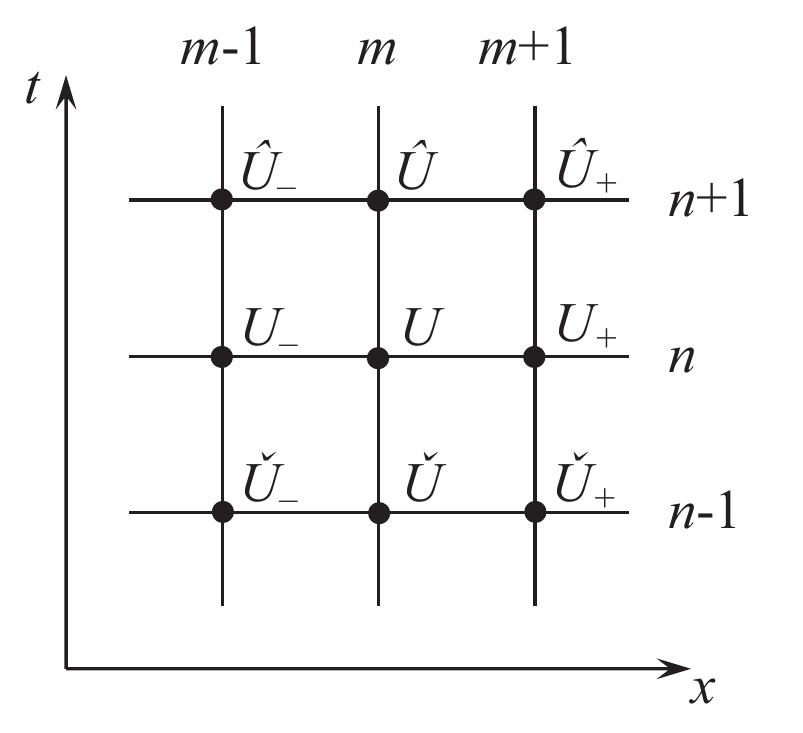} \caption{ \label{fig:stenc} The numerical mesh and the stencil centered at $(t_n, x_m)$.
}
\end{center}
\end{figure}

\medskip For the second-order PDE \eqref{eq:PDE2}, within a general nine-point stencil centered around $(t_n, x_m)$ (cf. Figure \ref{fig:stenc}), it is convenient to use the following index-free notation for the mesh points adjacent to the stencil center, and the approximate field values therein:
\begin{equation}\label{eq:hats}
\begin{array}{lll}
t_{n-1} \equiv \check{t}\,, & t_{n} \equiv t\,, & t_{n+1} \equiv \hat{t}\,,  \\[2ex]
x_{m-1} \equiv x_-\,, & x_m \equiv x\,, & x_{m+1} \equiv x_+\,,  \\[2ex]
U^{n-1}_{m-1} \equiv \check{U}_-\,,  &
U^{n-1}_{m} \equiv \check{U}\,, &
U^{n-1}_{m+1} \equiv \check{U}_+\,, \\[2ex]
U^{n}_{m-1} \equiv U_-\,,  &
U^{n}_{m} \equiv U\,, &
U^{n}_{m+1} \equiv U_+\,,  \\[2ex]
U^{n+1}_{m-1} \equiv \hat{U}_-\,,  &
U^{n+1}_{m} \equiv \hat{U}\,, &
U^{n+1}_{m+1} \equiv \hat{U}_+\,.
\earr
\end{equation}


%

The shift operators in the finite-difference space are defined as
\begin{equation}
\underset{\pm h}{S}: m \mapsto m \pm 1, \qquad
\underset{\pm \tau}{S}: n \mapsto n \pm 1, \qquad
\end{equation}
\begin{equation}
\underset{\pm h}{S}^k: m \mapsto m \pm k, \qquad
\underset{\pm \tau}{S}^k: n \mapsto n \pm k, \qquad
k = 0, 1, 2, \ldots \,,
\end{equation}
and act on a mesh quantities $x \equiv x_m$, $t\equiv t_n$,  $U^{n}_{m} \equiv U$ as follows:
\[
\SHP(x)=x_+,\quad \STP(t)=\hat{t},\quad \SHP(U)=U_+,\quad \underset{-h}{S}\,\underset{+\tau}{S}(U)=\hat{U}_-,
\]
etc., in a commutative fashion. In general, for a function of
finite-difference variables $t_n, x_m, U^n_m$, one has
\[
\underset{+h}{S}f(t_n, x_m, U^n_m) = f(t_n, x_{m+1}, U^n_{m+1}) \equiv f^+,
\quad
\underset{-h}{S}f(t_n, x_m, U^n_m) = f(t_n, x_{m-1}, U^n_{m-1}) \equiv f^-,
\]
\[
\underset{+\tau}{S}f(t_n, x_m, U^n_m) = f(t_{n+1}, x_m, U^{n+1}_m) \equiv \hat{f},
\quad
\underset{-\tau}{S}f(t_n, x_m, U^n_m) = f(t_{n-1}, x_m, U^{n-1}_m) \equiv \check{f}.
\]
The corresponding finite-difference operators are given by
\begin{eqnarray}
\DHP = \frac{\SHP - 1}{h_m}, \quad \DHM = \frac{1 - \SHM}{h_{m-1}}, \quad \DTP = \frac{\STP - 1}{\tau_n}, \quad \DTM = \frac{1 - \STM}{\tau_{n-1}}.
\end{eqnarray}
Using the difference and shift operators, one can write the discrete analogs of partial derivatives computed for the mesh quantity $U_{m}^{n}$ at various mesh points:
\begin{equation}\label{eq:discr:dxdt}
\begin{array}{lll}
U_x = \dfrac{U_{m+1}^n - U_m^n}{h_m} = \DHP(U), & U_{\bar{x}} = \dfrac{U^n_m - U_{m-1}^n}{h_{m-1}} = \DHM(U),  \\[2ex]
U_t = \dfrac{U^{n+1}_m - U^n_m}{\tau_{n}} = \DTP(U), & \check{U}_t=U_{\check{t}} = \dfrac{U^n_m - U^{n-1}_m}{\tau_{n-1}} = \DTM(U),\\[2ex]

U_{x\bar{x}} = \DHP \DHM(U), & U_{t\check{t}} = \DTP \DTM(U).
\earr
\end{equation}
Some shifted versions of the above first-order partial differences are given by
\beq
\barr
\hat{U}_x = \STP(U_x), \quad \hat{U}_{\bar{x}} = \STP(U_{\bar{x}}), \quad \check{U}_x = \STM(U_x), \quad \check{U}_{\bar{x}} = \STM(U_{\bar{x}}), \\[2ex]
U_t^+ = \SHP(U_t) , \quad \check{U}_t^+ = \SHP(\check{U}_t).
\earr
\eeq
For a general mesh \eqref{eq:mesh}, the discrete partial differences \eqref{eq:discr:dxdt} provide first-order approximations of the partial derivatives of the corresponding sufficiently smooth  continuous field $u(x,t)$:
\[
u_x(x_{m}, t_n)\simeq U_x,\quad u_{tt}(x_{m}, t_n)\simeq  U_{t\check{t}},
\]
etc. In the case of a uniform mesh, one has
\begin{equation} \label{regLattice}
h_{m+k} = h = \text{const}, \qquad
\tau_{n+l} = \tau = \text{const},
\qquad k, l \in \mathbb{Z},
\end{equation}
and the central second differences have a simpler form
\begin{equation}
U_{x\bar{x}} = \frac{U_{m+1}^n - 2 U^n_m + U_{m-1}^n}{h^2}, \qquad U_{t\check{t}} = \frac{U^{n+1}_m - 2 U^n_m + U_m^{n-1}}{\tau^2},
\end{equation}
providing second-order approximations of the second derivatives $u_{xx}$ and $u_{tt}$ at the mesh point $(x_{m}, t_n)$.

We note that the finite-difference operators in the same direction ($t$ or $x$) commute for general uniform (including moving) meshes:
\[
\DHP\DHM=\DHM\DHP, \quad \DTP\DTM=\DTM\DTP;
\]
moreover, on a stationary (uniform or non-uniform) mesh \eqref{eq:mesh:stationary}, difference operators in different directions also commute, for example, $\DTP\DHP=\DHP\DTP$.

\subsection{Invariance and conservation laws for PDEs}\label{sec:PDE:invar:cls}

For a general PDE system \eqref{eq:PDEgen}, a Lie algebra of linear differential operators
\begin{equation}\label{eq:Xk}
  \sg{X} =
  \xi^i(x,u) \, \frac{\partial}{\partial x^i}
  + \eta^\mu(x,u) \, \frac{\partial}{\partial u^\mu}
\end{equation}
corresponds to a Lie group of point transformations
\beq\label{eq:groupk}
\barrcl
(x^*)^i&=&f^i(x,u;\eps)=x^i + \eps \xi^i(x,u) + O(\eps^2),\quad i=1,\ldots ,p,\\[2ex]
(u^*)^\mu&=&g^\mu(x,u;\eps)=u^\mu + \eps \eta^\mu(x,u) + O(\eps^2),\quad \mu=1,\ldots ,q,
\earr
\eeq
of the $(p+q)-$dimensional space $(x, u)$ \cite{Ovsiannikov, OlverBk, BCAbook}. (In \eqref{eq:Xk} and below, summation in repeated indices is assumed.) In particular, an $s-$dimensional Lie algebra of generators  \eqref{eq:Xk} corresponds to an $s-$parameter Lie group $G_s$ of point transformations \eqref{eq:groupk}.

The $r^{\rm th}$ prolongation infinitesimal generator \eqref{eq:Xk} of a point transformation group \eqref{eq:groupk} provides the rules for transformation of derivatives of the dependent variables in the corresponding jet space $\mathcal{J}^r(x|u)$. In particular, the $r^{\rm th}$ prolongation of the evolutionary infinitesimal generator \eqref{eq:Xk} is given by
\beq\label{eq:ch1:sec11:prolongationXk}
{\rm{pr}}^{(r)}\,\sg{X}  \equiv \sg{X}^{(r)} = \sg{X}+ \eta^{(1)\,\mu}_{i}(x,u,\partial u) \dfrac{\partial }{\partial u^\mu_{i}} +\cdots + \eta^{(r)\,\mu}_{i_1\ldots i_r}(x,u,\partial u,\ldots,\partial^r u)\dfrac{\partial }{\partial
u^\mu_{{i_1}\ldots {i_r}}},
\eeq
where the higher-order infinitesimals are given by
\beq\label{eq:ch1:sec11:higher_etas}
\barrcl
\eta^{(1)\,\mu}_{i} &=& \sg{D}_{i}\eta^\mu - (\sg{D}_{i}\xi^j)\,u^\mu_{j} \\[2ex]
\eta^{(\ell)\,\mu}_{i_1\ldots i_\ell} &=& \sg{D}_{i_\ell}\eta^{(\ell-1)\,\mu}_{i_1\ldots i_{k-1}} - (\sg{D}_{i_\ell}\xi^j)\,u^\mu_{i_1\ldots i_{\ell-1} j},
\earr
\eeq
where $i, i_j=1,\ldots,p$, $\mu=1,\ldots,q$, and $\ell=2,3,\ldots\,r$. (For details, see, for example, \cite{Ovsiannikov, OlverBk, BCAbook}.) In \eqref{eq:ch1:sec11:higher_etas} and below, $\sg{D}_{i}F[u]$ denotes
the total derivative of a differential function $F[u]$ by $x_i$,
\beq\label{eq:ch1:sec11:tot_der}
\sg{D}_{i} F[u] =
\frac{\partial F[u]}{\partial x^i} + u^\mu_{i} \frac{\partial F[u]}{\partial u^\mu}
+ u^\mu_{i i_1}\frac{\partial F[u]}{\partial u^\mu_{i_1}}+\ldots,\quad i=1,\ldots ,p,
\eeq
and the notation $u^\mu_{i_1\ldots i_j}$ is used for partial derivatives of dependent variables:
\[
u^\mu_{i_1\ldots i_j} \equiv \dfrac{\partial^j \,u^\mu(x)}{\partial x^{i_1}\ldots \partial x^{i_j}}.
\]
For a single PDE $G[u]=0$ \eqref{eq:PDE2} with two independent variables $(x,t)$, the generator \eqref{eq:Xk} of the point transformation and its respective components may be denoted by
\begin{equation}\label{eq:Xk:2}
  \sg{X} = \xi^x(x,t,u) \, \frac{\partial}{\partial x} + \xi^t(x,t,u) \, \frac{\partial}{\partial t}
  + \eta(x,t,u) \, \frac{\partial}{\partial u}.
\end{equation}

\bigskip\noindent\textbf{Invariant functions and symmetries of PDEs.} A differential function $Q[u]$ defined on a jet space $\mathcal{J}^r(x|u)$ is \emph{invariant} with respect to the point transformation generated by \eqref{eq:Xk} (or equivalently, \eqref{point:eval}) if
\beq\label{dif:inv}
({\rm{pr}}^{(r)}\,\sg{{X}}) \, Q[u]\equiv 0,
\eeq
holding identically, i.e., for an arbitrary $u(x)$. (Such expressions $Q[u]$ are also called \emph{differential invariants}.)

For a PDE system $\Delta$ \eqref{eq:PDEgen} of order $k$, the notion of invariance is somewhat different; in particular, to be invariant with respect to a given point transformation (which is then called its symmetry), a PDE system \eqref{eq:PDEgen} has to satisfy the conditions
\beq\label{dif:inv:PDE}
({\rm{pr}}^{(k)}\,\sg{{X}}) \, R^\sigma[u]\Big|_\Delta =0, \quad
\sigma=1,\ldots ,K,
\eeq
where $(\cdot)\Big|_\Delta$ means that a quantity is computed on solutions of \eqref{eq:PDEgen}, that is, with some leading derivatives of the PDEs \eqref{eq:PDEgen} (and their differential consequences, as required) solved for in terms of other variables and derivatives, and substituted into the invariance condition \eqref{dif:inv:PDE} \cite{OlverBk, BCAbook}. In particular, if the PDEs \eqref{eq:PDEgen} satisfy the local solvability and maximal rank conditions, and $\sg{{X}}$ defines a point symmetry, then by Hadamard's lemma,  for each PDE $R^\sigma[u]$ of the system, the quantity $({\rm{pr}}^{(k)}\,\sg{{X}}) \, R^\sigma[u]$ is given by a linear combination of the PDEs $R^\sigma[u]$ and their differential consequences. When a PDE system is invariant with respect to an $s$-parameter Lie group $G_s$, the invariance condition \eqref{dif:inv:PDE} has to hold for each corresponding infinitesimal generator $\sg{X}=\sg{X}_j$, $j=1,\ldots, s$.

\medskip If a point transformation generator $\sg{{X}}$ \eqref{eq:Xk} defines a Lie point symmetry of a PDE system \eqref{eq:PDEgen}, it is often useful to consider the characteristic (evolutionary) form of $\sg{{X}}$, given by
\beq\label{point:eval}
\sg{\hat{X}} = \hat{\eta}^\mu[u] \,\dfrac{\partial}{\partial u^\mu} \equiv \hat{\eta}^\mu (x,u,\partial u,\ldots,\partial^s u)\,\dfrac{\partial}{\partial u^\mu},
\eeq
where
\[
\hat{\eta}^\mu[u]\equiv \hat{\eta}^\mu(x,u,\partial u)  = \eta^\mu(x,u) - u^\mu_i(x)\xi^i(x,u)
\]
are the evolutionary infinitesimal components. The global action of the transformation defined by \eqref{point:eval} is
\beq\label{eq:ev:groupk}
\barrcl
(x^*)^i&=&x^i,\quad i=1,\ldots ,p,\\[2ex]
(u^*)^\mu &=& u^\mu + \eps \hat{\eta}^\mu[u] + O(\eps^2), \quad \mu=1,\ldots,q.
\earr
\eeq
Prolongations ${\rm{pr}}^{(r)}\,\sg{\hat{X}}$ of evolutionary generators \eqref{point:eval} are computed in the same way as \eqref{eq:ch1:sec11:higher_etas}, assuming $\xi^i=0$, and using the extended dependence of $\hat{\eta}$. The $r^{\rm th}$ prolongation of the evolutionary infinitesimal generator \eqref{point:eval} acts on a differential function $Q[u]$ as
\beq\label{eq:prol:act:df}
({\rm{pr}}^{(r)}\,\sg{\hat{X}})\, Q[u] = ({\rm{pr}}^{(r)}\,\sg{{X}} + \xi^i\sg{D}_i)\, Q[u].
\eeq
Since on solutions of the PDE system $\Delta$ \eqref{eq:PDEgen}, all $R^\sigma[u]=\sg{D}_i R^\sigma[u]=0$, it follows that in the PDE system invariance condition \eqref{dif:inv:PDE}, one may replace $\sg{{X}}$ with ${\rm{pr}}^{(r)}\,\sg{\hat{X}}$.

We also note that in addition to the point transformations in the evolutionary form \eqref{point:eval}, one can consider more general higher-order transformations (sometimes called Lie-B\"{a}cklund transformations). These are given by \eqref{point:eval}, \eqref{eq:ev:groupk} with a generalized component dependence
\beq\label{point:eval:ho}
\hat{\eta}^\mu[u] = \hat{\eta}^\mu (x,u,\partial u,\ldots,\partial^s u),
\eeq
involving derivatives of $u$ up to some finite order $s>0$. The prolonged higher-order transformation generator still has the form similar to \eqref{eq:ch1:sec11:prolongationXk}, and the prolongation components are computed using formulas \eqref{eq:ch1:sec11:higher_etas} with $\xi^i=0$, applied to \eqref{point:eval:ho}. Further generalizations include, for example, nonlocal transformations, where the components $\hat{\eta}^\mu[u]$ essentially depend on nonlocal (e.g., potential) variables \cite{BCAbook}.

\bigskip\noindent\textbf{Conservation laws of PDEs.} For a PDE system $\Delta$ \eqref{eq:PDEgen}, a \emph{local conservation law} is given by a divergence expression
\begin{equation} \label{eq:defCL:Intro}
\sg{D}_i \Phi^i[u] =0,
\end{equation}
where differential functions $\Phi^i[u]$ are the conservation law fluxes (see, e.g., \cite{OlverBk, BCAbook}). For a single PDE $G[u]=0$ \eqref{eq:PDE2} with two independent variables $x,t$, the local conservation law takes the form
\beq\label{eq:defCL:2}
\sg{D}_t \Theta[u] +\sg{D}_x \Phi[u] =0,
\eeq
with the conserved density $\Theta[u]$. Globally, for $x\in[a,b]$, the local conservation law \eqref{eq:defCL:2} describes the evolution of the global quantity
\begin{equation} \label{eq:defConsQuant:Intro}
{Q}=\mathop{\int}_a^b \Theta[u]  \;dx,\qquad  \frac{d {Q}}{d t}=-\Phi[u]\Big|_a^b\,,
\end{equation}
in terms of the fluxes through the domain boundary; similar global forms hold in multi-dimensions \cite{BCAbook}. If the flux $\Phi$ vanishes on the domain boundary or at infinity, as well as in the periodic case, $Q$ defines a global conserved quantity: ${d Q}/{d t}=0.$  The local conservation law density and flux are not unique; they are defined up to adding a \emph{trivial conservation law} \cite{OlverBk, BCAbook, anco2018different}. Trivial conservation laws \eqref{eq:defCL:2} can be made of components of a total curl (e.g., $\sg{D}_t\sg{D}_x Z[u]-\sg{D}_x\sg{D}_t Z[u]=0$ for any $Z[u]$), and/or involve fluxes and density that vanish on solutions of the given system $\Delta$ \eqref{eq:PDEgen}. In general, for a given model, one seeks independent equivalence classes of local conservation laws \eqref{eq:defCL:2}, written in simplest forms modulo adding trivial conservation laws.

If a given PDE system is totally nondegenerate, it follows from the Hadamard lemma that every nontrivial local conservation law \eqref{eq:defCL:Intro} (up to equivalence) can be written in a \emph{characteristic form} \cite{OlverBk, BCAbook}
\beq\label{eq:CL:charform}
\sg{D}_i \tilde{\Phi}^i[u] = \Lambda_\sigma[u]R^\sigma[u]=0,
\eeq
where $\{\Lambda_\sigma[u]\}_{\sigma=1}^K$ are the conservation law multipliers (characteristics), not all zero, and $\sg{D}_i (\Phi^i[u]-\tilde{\Phi}^i[u])\equiv 0$ is a trivial conservation law. In practice, one is interested in computing the full set of linearly independent, nontrivial local conservation laws of a given PDE system, i.e., in obtaining one representative of each conservation law equivalence class.

When the PDE system $\Delta$ \eqref{eq:PDEgen} is \emph{variational}, that is, when all equations arise as Euler-Lagrange equations of some action functional, the first Noether's theorem establishes a one-to-one correspondence between conservation law multipliers and variational symmetries of the given system. A PDE system is variational as it stands when its linearization operator (Fr\'{e}chet derivative) is self-adjoint. Few PDEs and PDE systems arising in practical applications turn out to be variational. Moreover, the property of a PDE system being variational depends on the choice of variables and representation of a PDE system \cite{BCApap}. The question of determination whether or not a given system is equivalent to a variational one generally remains open \cite{OlverBk, BCAbook}. For non-variational systems, it is well-known that sets of local conservation laws and local symmetries can be quite different. In particular, there may be more point symmetries than conservation laws, and vice versa, it is easy to construct examples of PDEs with at least one conservation law and no point symmetries.

The \emph{direct method} of seeking conservation laws provides a systematic and efficient approach to local conservation law construction, applicable to general (that is, not only variational) PDE systems \eqref{eq:PDEgen} \cite{alonso1979noether, vinogradov1984local, OlverBk, Anco1997,Anco2002part1,BCAbook}. The method consists in the application of Euler differential operators
to seek sets of local {multipliers} $\{M_\sigma[u]\}_{\sigma=1}^N$ yielding local conservation laws in characteristic form \eqref{eq:CL:charform}. In particular, a differential function $M_\sigma[u]R^\sigma[u]$ is a divergence expression if and only if it is identically annihilated by all Euler operators
\beq\label{EulerOp}
\sg{E}_{u^j} \equiv \frac{\partial }{\partial u^j} - \sg{D}_i \frac{\partial }{\partial u_i^j
} + \cdots + ( - 1)^s\sg{D}_{i_1 } \ldots \sg{D}_{i_s } \frac{\partial }{\partial u_{i_1 \ldots i_s }^j } + \cdots,\qquad j=1,\ldots, q,
\eeq
for an arbitrary vector function $u(x)$. Then on the solution set  of \eqref{eq:PDEgen}, a local conservation law \eqref{eq:CL:charform} holds.
The determining equations
\beq\label{EulerOp:directM}
\sg{E}_{u^j} (M_\sigma[u]R^\sigma[u])=0
\eeq
are linear overdetermined PDEs for the unknown multipliers; they can be systematically solved by hand or using symbolic software, and the multipliers in turn yield conservation law density and/or fluxes. Multiple examples of the use of the direct method can be found, for example, in Ref.~\cite{BCAbook} and references therein.

The direct method and related methods of flux computation have been implemented in the symbolic package \verb|GeM| for \verb|Maple| (see \cite{GeM,C_flux}). Other symbolic software packages for conservation law computations exist; see, e.g., \cite{wolf2002comparison,C_flux}.

\subsection{Invariance and conservation laws for difference equations}

For partial difference equations, in a way similar to PDEs, a \emph{difference invariant} of the Lie group $G$ with an infinitesimal generator  $\sg{X}$ \eqref{eq:Xk} is any difference expression
\begin{equation}\label{hinv}
  I = I(\mathbf{t}, \mathbf{x}, \mathbf{U}),
\end{equation}
defined on the stencil and satisfying the identity
\begin{equation} \label{hinvCond}
  (\underset{h}{\rm{pr}}\, \sg{X})\, I \equiv 0,
\end{equation}
where $\underset{h}{\rm{pr}}\, \sg{X}$ defines the prolongation of $\sg{X}$ on the finite-difference stencil
$(\mathbf{t}, \mathbf{x}, \mathbf{U})$, and includes components for $x$, $U$, for discrete analogs of partial derivatives \eqref{eq:discr:dxdt}, as well as components for local mesh steps \cite{VDGroups,VDBook}. If $I(\mathbf{t}, \mathbf{x}, \mathbf{U})$ is an invariant with respect to an $s$-parameter Lie group $G_s$, the condition \eqref{hinvCond} has to hold for each corresponding infinitesimal generator $\sg{X}_j$, $j=1,\ldots, s$.


In a way parallel to the PDE invariance, a difference equation \eqref{eq:diff:F}
is invariant with respect to a Lie group of point transformations with a generator $\sg{X}$ if
\begin{equation}\label{eq:diff:H:invcond}
  (\underset{h}{\text{pr}}\, \sg{X})\, F |_{[F]} = 0, \qquad j = 1, \dots, s,
\end{equation}
where $[F]$ denotes the condition $F=0$ \eqref{eq:diff:F} holding at a given and adjacent stencils, as well as \emph{difference consequences} of such equations. For example, if \eqref{eq:diff:F} holds, then difference consequences
\[
\DHP F=0, \quad \DHM \, F=0, \quad \DTP\, F=0, \quad \DTM \, F=0, \quad \DHP \DHM \, F=0,\quad \ldots,
\]
are naturally required to hold, and are used in the computation of the invariance conditions \eqref{eq:diff:H:invcond}. For a finite-difference scheme \eqref{schemeGeneral} to be invariant as a whole, it should satisfy the conditions
\begin{equation} \label{hSymsSys}
(\underset{h}{\text{pr}}\, \sg{X})\, F |_{[\Delta_h]} = 0,  \quad
(\underset{h}{\text{pr}}\, \sg{X})\, \Omega^k |_{[\Delta_h]} = 0,  \quad k = 1, 2.
\end{equation}
where $[\Delta_h]$ denotes the scheme equations \eqref{schemeGeneral} with all their required difference consequences. In particular, if the finite-difference scheme is composed of difference invariants of the form \eqref{hinv}, then such a scheme would satisfy the above invariance conditions \eqref{hSymsSys}. [The converse is generally not true, but the use of difference invariants provides a straightforward systematic method to construct invariant schemes.] A finite-difference scheme \eqref{schemeGeneral} is invariant with respect to an $s$-parameter Lie group $G_s$ when the invariance conditions \eqref{hSymsSys} hold for each of the $s$ linearly independent infinitesimal generators $\sg{X}=\sg{X}_j$, $j=1,\ldots, s$.

\bigskip\noindent\textbf{Local conservation laws of difference equations.} A local \emph{difference conservation law}
is a divergence expression
\begin{equation} \label{hdiv}
  \mathcal{K}:= \DTM\Theta(\mathbf{t}, \mathbf{x}, \mathbf{U})
  + \DHM\Phi(\mathbf{t}, \mathbf{x}, \mathbf{U}) =0,
\end{equation}
holding on solutions of a given finite-difference scheme $\Delta_h$ \eqref{schemeGeneral}.
The notion of triviality for difference conservation laws is the same as for the continuous case: a trivial difference conservation law \eqref{hdiv} has a part that holds as a difference identity for any $\mathbf{U}$, and/or a part whose density $\Theta$ and/or flux $\Phi$ vanish on solutions of the scheme \eqref{schemeGeneral}.

Similarly to the continuous case, the discrete conservation laws have a direct global interpretation. Let $\mathcal{K}^n_m$ denote the expression \eqref{hdiv} written on the numerical stencil centered at $(x_m, t_n)$ (see, e.g.,  Fig.~\ref{fig:stenc}). Then multiplying the difference expression by $h_{m-1}$ and summing from $m=m_1$ to $m=m_1+\ell$, $\ell\geq 1$, one gets
\begin{equation} \label{difference:integral}
\DTM {Q}_h = \Phi^n_{m+\ell}-\Phi^n_{m-1}.
\end{equation}
In \eqref{difference:integral}, the quantity ${Q}_h$ denotes the ``total amount" of a physical quantity with linear density $\Theta$, located between the mesh nodes $m=m_1-1$ and $m=m_1+\ell$; $\DTM {Q}_h$ is its discrete rate of change from $t=t_{n-1}$ to $t_n$. Similarly to the continuous case, this rate of change of a global quantity ${Q}_h$ is determined by the flux values $\Phi^n_{m+\ell}$ and $\Phi^n_{m-1}$ at the boundaries of the summation domain. The formula \eqref{difference:integral} is thus fully parallel to the continuum version \eqref{eq:defConsQuant:Intro} of a global conservation law.

\medskip
Again similarly to PDEs, for difference equations, conservation laws can be sought in a characteristic form. In particular, for a finite-difference equation $F=0$ \eqref{eq:diff:F}, the characteristic form is given by, for example,
\begin{equation}\label{eq:CL:discr}
  \DTM\Theta(\mathbf{t}, \mathbf{x}, \mathbf{U}) + \DHM\Phi(\mathbf{t}, \mathbf{x}, \mathbf{U}) = \Lambda F(\mathbf{t}, \mathbf{x}, \mathbf{U})  ,
\end{equation}
where $\Lambda = \Lambda(\mathbf{t}, \mathbf{x}, \mathbf{U})$ is the corresponding characteristic (also referred to as an \emph{integrating factor}, especially in the context of ODEs). A known characteristic of a difference conservation law \eqref{hdiv} corresponds to an equivalence class of density-flux pairs $(\Theta,\Phi)$ (up to adding trivial conservation laws).

The \emph{direct method} to seek difference conservation laws \eqref{eq:CL:discr} is parallel to that for PDEs. For example, for the case of a single PDE \eqref{eq:PDE2} on $u(t,x)$ and its finite-difference approximation \eqref{schemeGeneral}, the corresponding \emph{difference Euler operator} written around the stencil point $U$ of the uniform mesh \eqref{regLattice} has the form \cite{VDBook}
\begin{equation}\label{EulerOpH}
  \mathcal{E}_U = \sum_{k=-\infty}^{\infty} \, \sum_{l=-\infty}^{\infty} \, {\STM}^k {\SHM}^l \left(
    \frac{\partial}{\partial U_{m+l}^{n+k}}
  \right).
\end{equation}
As in the continuous case, the difference Euler operator \eqref{EulerOpH} annihilates all difference-type divergence expression \eqref{hdiv} on the mesh \eqref{regLattice}:
\begin{equation}
\mathcal{E}_U \mathcal{K}|_{\eqref{regLattice}} \equiv 0,
\end{equation}
which holds for an \emph{arbitrary} mesh quantity $\mathbf{U}$, not only the solutions of the given finite-difference equation \eqref{eq:diff:F}.  Consequently, for a given difference equation \eqref{eq:diff:F} holding on the lattice \eqref{regLattice}, its integrating factors $\Lambda(\mathbf{t}, \mathbf{x}, \mathbf{U})$ are found from the determining equations
\begin{equation}
\mathcal{E}_U (\Lambda F)|_{(\ref{regLattice})} \equiv 0,
\end{equation}
holding identically for an arbitrary $\mathbf{U}$.

\medskip

As opposed to the differential case, in which the differential equation is always given, the difference approximation $F$ \eqref{eq:diff:F} is usually not known in advance. If the exact form of $F$ is not known, then one can consider $F$ as a difference expression with unknown arbitrary coefficients. In order to find the values of these coefficients, one can use, for example, symmetry invariance conditions \eqref{hSymsSys}. In addition, because the scheme should approximate the differential problem, to further specify  the form of $F$ one can use the method of undetermined coefficients \cite{godunov1973raznostnye}.

\medskip

Consider the case when a difference scheme \eqref{schemeGeneral}, approximating a scalar PDE \eqref{eq:PDE2} with one dependent and two independent variables, is defined on a uniform orthogonal mesh \eqref{regLattice}. In this case, it is appropriate to use the Euler operator \eqref {EulerOpH}. Let us also assume that an $r-$ parameter transformation group $G_r$  generated by the operators $\sg{X}_1$, $\ldots$, $\sg{X}_r$ (of the form \eqref{eq:Xk:2}) leaves invariant the uniform orthogonal mesh \eqref{regLattice}.
This is equivalent to requiring \cite{VDBook}
\beq\label{latticeConstraints1}
\barr
  \DHP \DHM (\xi^x_k) = 0, \qquad \DTP \DTM (\xi^t_k) = 0, \qquad
  \underset{\pm h}{D}(\xi^t_k) = -\underset{\pm \tau}{D}(\xi^x_k),
  \qquad k = 1, \dots, r.
\earr
\eeq
Then the system $\Delta_h$ \eqref{schemeGeneral} is simplified, and together with conditions \eqref{latticeConstraints1}, takes the form
\begin{equation}\label{detSys0}
\barr
\Delta_h:~ \begin{cases}
  F(\mathbf{t}, \mathbf{x}, \mathbf{U}) = 0, \\
  h_{m+i} = h, \quad i \in \mathbb{Z}, \\
  \tau_{n+j} = \tau, \quad j \in \mathbb{Z},
\end{cases} \\[5ex]
(\underset{h}{\text{pr}} \,\sg{X}_k) F |_{[\Delta_h]} = 0, \quad k = 1, \dots, r,\\[2ex]
\DHP \DHM (\xi^x_k) = 0, \quad \DTP \DTM (\xi^t_k) = 0, \quad
\underset{\pm h}{D}(\xi^t_k) = -\underset{\pm \tau}{D}(\xi^x_k),\quad k = 1, \dots, r.
\earr
\end{equation}
When a concrete form of the difference approximation $F(\mathbf{t}, \mathbf{x}, \mathbf{U})$ of a PDE model is not specified, one may seek both conservation law multipliers $\Lambda_j(\mathbf{t}, \mathbf{x}, \mathbf{U})$ and the form of $F$. In this case, the system \eqref{detSys0} is appended by the equations
\begin{equation}\label{detSys0Eu}
\mathcal{E}_U (\Lambda_j F)|_{\eqref{regLattice}} \equiv 0, \qquad j = 1, 2, \dots\,.
\end{equation}

%



\section{Example 1: an invariant scheme with conservation laws for the  linear wave equation}\label{sec:linW}

\subsection{Local symmetries and conservation laws}\label{sec:linW:S:CL}

As a basic example, we consider the linear homogeneous wave equation
\begin{equation} \label{W}
W = u_{tt} - u_{xx} = 0,
\end{equation}
(cf. \eqref{eq:lin:Wave:intro}). It admits an infinite-parameter point symmetry group generated by the infinitesimal operators \eqref{eq:Xk:2} with
\beq\label{W:symGen:eta} 
\eta=C_1u + \alpha(x,t),
\eeq
where $C_1=\const$, and $\alpha(x,t)$ is an arbitrary solution of the PDE \eqref{W}. The latter corresponds to the infinite-parameter symmetry group
\[
u(x,t)\to u(x,t)+\alpha(x,t)
\]
admitted by any linear homogeneous equation; the generator for these point symmetries is given by
\begin{equation}\label{eq:W:sym:alpha}
\sg{X}_{\alpha} = \alpha(x,t)\frac{\partial}{\partial u}.
\end{equation}
The symmetry components for the linear wave equation $\xi^x$ and $\xi^t$ are independent of $u$. Both of them are given by arbitrary solutions of the PDE \eqref{W}, satisfying a mutual relation
\beq\label{W:symGen:Xis:Eq}
\xi^x_x=\xi^t_t,\quad \xi^x_t=\xi^t_x.
\eeq
The explicit form of these components can be written explicitly, for example, using the method of characteristics, as
\[\xi^x=f(x+t)+g(x-t),\quad\xi^t=f(x+t)-g(x-t)\]
in terms of a pair of smooth arbitrary functions $f(z)$, $g(z)$. The general point symmetry generator  of an infinite-parameter Lie group for the linear wave equation \eqref{W} is thus given by
\begin{equation}\label{eq:W:sym:gen}
\sg{X} = \xi^x\frac{\partial}{\partial x}+\xi^t\frac{\partial}{\partial t}+  \eta\frac{\partial}{\partial u},
\end{equation}
with components satisfying \eqref{W:symGen:eta}, \eqref{W:symGen:Xis:Eq}. In particular, it includes
the basic 
symmetries
\begin{equation}\label{W:sym:geom}
\barr
\sg{X}_1 = \dfrac{\partial}{\partial u}, \quad
\sg{X}_2 = \dfrac{\partial}{\partial x}, \quad
\sg{X}_3 = \dfrac{\partial}{\partial t}, \quad
\sg{X}_4 = t \, \dfrac{\partial}{\partial u},\quad
\sg{X}_5 = x \, \dfrac{\partial}{\partial u}, \\[2ex]
\qquad
\sg{X}_6 = x \, \dfrac{\partial}{\partial x} + t \, \dfrac{\partial}{\partial t},\quad \sg{X}_7 = x \, \dfrac{\partial}{\partial t} + t \, \dfrac{\partial}{\partial x},\quad
\sg{X}_8 = u \, \dfrac{\partial}{\partial u},
\earr
\end{equation}
where $\sg{X}_1$, $\sg{X}_2$ are spatial translations, $\sg{X}_3$ is a time translation, $\sg{X}_4$ is the Galilei symmetry,  $\sg{X}_5$ is the uniform stretching transformation,  $\sg{X}_6$ and $\sg{X}_8$ are scalings of independent and dependent variables, and $\sg{X}_7$ is a Lorentz boost. [We note that the full set of (higher-order) symmetries, variational symmetries, and local conservation laws of the linear wave equation \eqref{W} is described in Ref.~\cite{WavePC}.]

\medskip
Since the wave equation \eqref{W} is the Euler-Lagrange equation for the Lagrangian density
\begin{equation}\label{WL}
  L=L[u] = \dfrac{1}{2}(u_x^2 - u_t^2),
\end{equation}
the conservation laws of \eqref{W} can be computed using the first Noether's theorem \cite{Noether1918, OlverBk} and the variational symmetries of \eqref{W}. In particular, seeking conservation laws corresponding to the symmetries \eqref{W:sym:geom}, one observes that the action integral
\[
\displaystyle
S=\int_{t_0}^{t_1}\!\int_{x_0}^{x_1}L\,dx\,dt
\]
is invariant with respect to the symmetries $\sg{X}_1$,\ldots, $\sg{X}_7$, hence the conservation law multipliers are given by the corresponding evolutionary symmetry components: $M_1=1$, $M_2=u_x$, $M_3=u_t$, $M_4=t$, $M_5=x$, $M_6=xu_x+tu_t$, $M_7=tu_x+xu_t$. In particular, symmetries
$\sg{X}_1$, $\sg{X}_2$, $\sg{X}_3$,
$\sg{X}_6$ and $\sg{X}_7$
preserve the Lagrangian \eqref{WL} exactly, whereas the symmetries $\sg{X}_4$ and $\sg{X}_5$
preserve it to within a divergence:
\begin{subequations}\label{LinW:Symms:action:on:Lagr}
\begin{eqnarray}
  {\rm pr}\,\hat{X}_4 (L) + L (D_x\xi^x_4 + D_t \xi_4^t) &=& -D_t(u), \\[2ex]
  {\rm pr}\,\hat{X}_5 (L) + L (D_x\xi^x_5 + D_t \xi_5^t) &=& D_x(u).
\end{eqnarray}
\end{subequations}
The scaling symmetry $\sg{X}_8$ does not preserve the action, and hence is non-variational, with no corresponding conservation law. In summary, the conservation law multipliers and divergence expressions corresponding to the geometrical symmetries \eqref{W:sym:geom} are given by
\begin{subequations}\label{CLsW}
\begin{eqnarray}
M_1=1, && D_t\left( u_t \right) - D_x\left( u_x \right) = 0, \label{CLsW:sub1} \\[2ex]
M_2=u_x, &&D_t\left( u_t u_x \right) - D_x\left( \frac{u_t^2 + u_x^2}{2} \right) = 0,\label{CLsW:sub2}\\[2ex]
M_3=u_t, &&D_t\left( \frac{u_t^2 + u_x^2}{2} \right) - D_x\left( u_t u_x \right) = 0, \label{CLsW:sub3}\\[2ex]
M_4=t, &&D_t\left(t u_t - u \right) - D_x\left( t u_x \right) = 0, \label{CLsW:sub4}\\[2ex]
M_5=x, &&D_t\left(x u_t \right) - D_x\left( x u_x -u\right) = 0, \label{CLsW:sub4x}\\[2ex]
M_6=xu_x+tu_t, &&D_t\left(x u_t u_x  +\dfrac{t}{2}(u_t^2+u_x^2) \right) - D_x\left( t  u_t u_x +\dfrac{x}{2}(u_t^2+u_x^2) \right) = 0, \label{CLsW:sub5}\\[2ex]
M_7=tu_x+xu_t, &&D_t\left(t u_t u_x  +\dfrac{x}{2}(u_t^2+u_x^2) \right) - D_x\left( x  u_t u_x +\dfrac{t}{2}(u_t^2+u_x^2) \right) = 0. \label{CLsW:sub6}
\end{eqnarray}
\end{subequations}
In particular, the first conservation law describes the local conservation of momentum, the third one -- the conservation of mechanical energy, and the fourth one -- the motion of the center of mass in the displacement direction $z$ (see, e.g., \cite{cheviakov2016one}, or \cite{OlverBk} p.~279):
\beq\label{cmass}
Z_c = Z_0 + V t,
\eeq
where $Z_c$ is the position of the center of mass, $Z_0$ is its initial position, and $V$ is the center of mass velocity. (The latter interpretation holds when $u_x=0$ on the domain boundaries. This is the case, for example, for small longitudinal oscillations of an ideal elastic rod with free ends. The formula \eqref{cmass} arises from the global form \eqref{eq:defConsQuant:Intro} of the conservation law \eqref{CLsW:sub4}.) The physical meaning of the conservation law \eqref{CLsW:sub4x} is evident for the wave equation describing small transverse displacements $u(t,x)$ of an oscillating string of length $\ell$, $0\leq x \leq \ell$. There, the conserved density $xu_t$ is the density of angular momentum about $x=0$. Similarly, equivalent conservation laws with conserved densities $(x-a)u_t$, for $a=\const$, $0\leq a \leq \ell$, describe the conservation of the total angular momentum about $x=0$.

It is straightforward to show that the linearity symmetries $\sg{X}_{\alpha}$ \eqref{eq:W:sym:alpha} with $\alpha(x,t)$ satisfying $\alpha_{tt}=\alpha_{xx}$ also preserve the Lagrangian density to within a divergence, and yield conservation laws
\begin{eqnarray}
M_\alpha=\alpha(x,t), && D_t\left( \alpha u_t -\alpha_t u \right) - D_x\left( \alpha u_x -\alpha_x u\right) = 0. \label{CLsW:subA}
\end{eqnarray}

\medskip
Alternatively to the use of Noether's theorem, and technically in a more straightforward way (see, e.g., \cite{BCA, BCAbook}), conservation laws of the linear wave equation \eqref{W}  can be derived using the direct method involving the characteristic form and multipliers (Section \ref{sec:PDE:invar:cls}). The multiplier determining equations \eqref{EulerOp:directM} for the current example take the form
\beq\label{EulerOp:directM:wave}
\sg{E}_{u} (M\, W)\equiv 0.
\eeq
Seeking, for example, \emph{first-order} multipliers (i.e., expressions that may involve first derivatives) $M[u]=M(x,t, u, u_x, u_t)$, one obtains the split determining equations
\beq\label{eq:linW:CL:Mult:DetEq}
M_u=0,\quad M_{tt}=M_{xx}, \quad M_{t\,u_t}=M_{x\,u_x},\quad M_{t\,u_x}=M_{x\,u_t}, \quad M_{u_t\,u_t}=M_{u_x\,u_x},
\eeq
which include as particular solutions all multipliers listed in \eqref{CLsW} and \eqref{CLsW:subA}. (We note that in order to describe the full set of local conservation laws of the linear wave equation \eqref{W}, it is technically preferable to use the ``brute force" approach~\cite{WavePC}.)

\subsection{An invariant conservation law-preserving discretization}\label{sec:linW:discr}

We now wish to construct an invariant finite-difference scheme for the PDE \eqref{W} on the uniform orthogonal mesh \eqref{regLattice}, requiring that this scheme possesses difference analogs of the conservation laws \eqref{CLsW}. Operators $\sg{X}_2$, \ldots, $\sg{X}_6$ and $\sg{X}_8$ in \eqref{W:sym:geom} preserve the mesh orthogonality and uniformity, since for these operators, all conditions  \eqref{latticeConstraints1}  are satisfied. We note that any transformation involving only changes in dependent variables, including the infinite set of linearity symmetries $\sg{X}_{\alpha}$ \eqref{eq:W:sym:alpha} and their particular instance, the translation operator $\sg{X}_1$ in \eqref{W:sym:geom}, automatically satisfy \eqref{latticeConstraints1}, and therefore do not modify the mesh. The Lorentz boost operator $\sg{X}_7$ fails to satisfy the conditions \eqref{latticeConstraints1}, and therefore fails to preserve the mesh properties (cf.~\cite{VDBook}).


\medskip

\begin{figure}[h]
\centering
\includegraphics[scale=0.75]{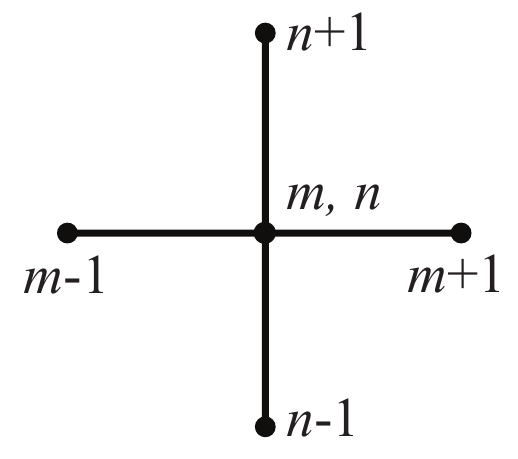}
\caption{The five-point cross-type stencil.}
\label{fig:5pt-stencil}
\end{figure}

The linear PDE \eqref{W} is rather simple, so it is natural to consider the basic explicit scheme on a five-point cross-type stencil (Fig.~\ref{fig:5pt-stencil}), given by
\begin{equation}\label{Wscheme}
\begin{cases}
  \mathcal{W} \equiv U_{t\check{t}} - U_{x\bar{x}} = 0, \\
  h_{m+k} = h = \text{const}, \quad k \in \mathbb{Z}, \\
  \tau_{n+l} = \tau = \text{const}, \quad l \in \mathbb{Z}.
\end{cases}
\end{equation}
in terms of the finite-difference approximations of derivatives \eqref{eq:discr:dxdt}. The difference equation $\mathcal{W}$ involves symmetric central second differences in space and time,
\[
 \mathcal{W} = \dfrac{1}{\tau^2}(\hat{U}+\check{U}-2U) - \dfrac{1}{h^2}({U}_+ + {U}_{-}-2U) = 0,
\]
and provides a second-order approximation of the wave equation \eqref{W} in terms of both $h$ and $\tau$.

The difference equation $\mathcal{W}$ in \eqref{Wscheme} can be written in a divergence form equivalent to \eqref{hdiv}, specifically,
\begin{equation}\label{W:CLFORM}
  \mathcal{W} = \DTM(U_t) - \DHM(U_x) = 0,
\end{equation}
which is a finite-difference analog of the first conservation law \eqref{CLsW:sub1}, with the multiplier $\Lambda=1$.
Also, since
\begin{equation}\label{eq:ClW4:Eu}
  \mathcal{E}_U (t \mathcal W) \equiv 0,
\end{equation}
it is evident that $\Lambda=t$ is a multiplier; in particular, it yields a divergence expression
\begin{equation}\label{eq:ClW4:form}
  t (U_{t\check{t}} - U_{x\bar{x}}) = \DTM(\hat{t} U_t - \hat{U}) - \DHM(t U_x) = 0,
\end{equation}
which is a finite-difference version of the conservation law \eqref{CLsW:sub4}. (In \eqref{eq:ClW4:Eu} and \eqref{eq:ClW4:form},  $t=t_n$ is the discrete time value at the center of the stencil.) By symmetry of $x$ and $t$, $\Lambda=x$ is also a valid conservation law multiplier of the difference wave equation \eqref{W:CLFORM}.

To find two finite-difference conservation laws corresponding to \eqref{CLsW:sub2}, \eqref{CLsW:sub3}, we seek integrating factors as mesh quantities defined on the same stencil, in general, as expressions
\begin{equation}
\Lambda_z = z_1 U^n_m + z_2 U^n_{m-1} + z_3 U^n_{m+1} + z_4 U^{n-1}_m + z_5 U^{n+1}_m,
\end{equation}
where $z_1, \dots, z_5$ are constant parameters. The conservation law determining equations yield
\beq
\barr
 \mathcal{E}_U \left( \Lambda_z \mathcal{W} \right) =&
\dfrac{z_4 + z_5}{\tau^2}\, (U^{n-2}_m + U^{n+2}_m)
- \dfrac{z_2 + z_3}{h^2} \, (U^n_{m-2} + U^n_{m+2}) \\[2ex]
& + \left( \dfrac{z_2 + z_3}{\tau^2} - \dfrac{z_4 + z_5}{h^2} \right) \,
(U^{n-1}_{m-1} + U^{n-1}_{m+1} + U^{n+1}_{m-1} + U^{n+1}_{m+1}) \\[2ex]
&+ 2 \left( \dfrac{2 z_1 - z_2 - z_3}{h^2} - \dfrac{2 z_1 - z_4 - z_5}{\tau^2} \right) U^n_m \\[2ex]
&+ 2 \left( \dfrac{z_2 + z_3 - z_1}{h^2} - \dfrac{z_2 + z_3}{\tau^2} \right) ( U^n_{m-1} + U^n_{m+1}) \\[2ex]
& + 2 \left( \dfrac{z_4 + z_5}{h^2} + \dfrac{z_1 - z_4 - z_5}{\tau^2} \right) ( U^{n-1}_m + U^{n+1}_m ) \equiv 0.
\earr
\eeq
Since the above expression must vanish identically, it follows that coefficients at all different mesh values of $U_a^b$ must vanish independently. Solving for $z_i$, we obtain
\begin{equation}
  z_1 = 0, \quad z_3 = -z_2, \quad z_5 = -z_4.
\end{equation}
It follows that admissible finite-difference conservation law multipliers have the form
\begin{equation}
  \Lambda_z = z_2 (U^n_{m-1} - U^n_{m+1}) + z_4 (U^{n-1}_m - U^{n+1}_m)
  \equiv -2 z_2 \, \frac{U_x + {U_{\bar{x}}}}{2} - 2 z_4 \, \frac{U_t + \check{U}_t}{2}.
\end{equation}
It is straightforward to observe that
\begin{equation}\label{eq:LinW:Lam23}
\Lambda_2 = \frac{U_x + U_{\bar{x}}}{2}, \qquad
\Lambda_3 = \frac{U_t + \check{U}_t}{2}
\end{equation}
approximate respectively the conservation laws multipliers $M_2 = u_x$ and $M_3 = u_t$ of the continuum conservation laws \eqref{CLsW:sub2} and \eqref{CLsW:sub3} of the PDE $W$ \eqref{W}.  It remains to write down the corresponding finite-difference divergence expressions. First, observe that in the continuum case, for the conservation law multiplier $M_2$, from \eqref{CLsW:sub2}, we have
\begin{equation} \label{QRrelDiff}
D_t(M_2 u_t) - M_2 W = D_x \left( \frac{u_t^2 + u_x^2}{2} \right).
\end{equation}
This relation carries over to the difference case:
\begin{equation}
\DTM\left(
    u_t \, \STP(\Lambda_2)
\right)
 - \Lambda_2 \, \mathcal{W} =
 \DHM\left(
 \frac{U_t\, U_t^+ + U_x^2}{2}
    \right),
\end{equation}
so the discrete conservation law corresponding to \eqref{CLsW:sub2} is given by
\begin{equation}
\DTM\left(
    U_t \, \frac{\hat{U}_x + \hat{U}_{\bar{x}}}{2}
    \right)
- \DHM\left(
 \frac{U_t\, U_t^+ + U_x^2}{2}
    \right) = 0.
\end{equation}
Due to the symmetry of  $x\leftrightarrow t$ in the model and the multipliers $\Lambda_2$, $\Lambda_3$, it is straightforward to write down the conservation law corresponding to $\Lambda_3$.

\medskip Further, we attempt to find discrete conservation law multipliers related to $M_6$ and $M_7$ \eqref{CLsW:sub5}, \eqref{CLsW:sub6}. As before, we seek discrete $\Lambda$ in terms of combinations of components of $t$ and $x$ and discrete partial differences defined on the five-point cross  stencil (Fig.~\ref{fig:5pt-stencil}). This can be done with lengthy computations using the method of undetermined coefficients \cite{godunov1973raznostnye}. Undetermined coefficients involve combinations of variables present in multipliers $M_6$ and $M_7$. The forms of these expressions must satisfy the determining equations \eqref{detSys0Eu}, and also in the continuum limit, yield the expressions of $M_6$ and $M_7$. With the help of computer algebra, it is rather straightforward to find the analog of $M_7$: the discrete multiplier is given by a simple expression
\[
\Lambda_7 =t\Lambda_2 + x \Lambda_3 = t \dfrac{U_x + U_{\bar{x}}}{2} +x \dfrac{U_t + \check{U}_t}{2};
\]
where $t=t_n$ and $x=x_m$ are the discrete time and space values computed in the center of the stencil, as in \eqref{eq:hats}. The density and flux can also be computed.

Conversely, through detailed computations, it is possible to show that the finite-difference analog of $M_6$ \emph{does not exist} on the five-point cross stencil. (Such an analog, however, exists, and is not unique, on the nine-point stencil; see Fig.~\ref{fig:9pt-stencil} in the section below. Hence the discrete analog of the continuum conservation law \eqref{CLsW:sub5},  can hold for more general numerical methods on a uniform orthogonal mesh, but not for the explicit scheme defined by \eqref{Wscheme}.)

Finally, we have a symmetry-invariant finite-difference scheme \eqref{Wscheme} with local conservation laws
\begin{subequations} \label{hWCLs}
\begin{eqnarray}
\Lambda_1 = 1, &&
\DTM(U_t) - \DHM\left( U_x \right) = 0, \\[2ex]
\Lambda_2 = \frac{U_x + U_{\bar{x}}}{2}, &&
\DTM\left(
    U_t \, \frac{\hat{U}_x + \hat{U}_{\bar{x}}}{2}
\right)
 - \DHM\left(
 \frac{U_t U_t^+ + U_x^2}{2}
    \right)
= 0, \label{hWCLs:b}\\[2ex]
\Lambda_3 = \frac{U_t + \check{U}_t}{2}, &&
\DTM\left(
    \frac{U_x \hat{U}_x + U_t^2}{2}
\right)
 - \DHM\left(
    U_x \, \frac{U_t^+ + \check{U}_t^+}{2}
\right) = 0, \label{hWCLs:c}\\[2ex]
\Lambda_4 = t, &&
\DTM(\hat{t} U_t - \hat{U}) - \DHM\left( t U_x \right) = 0,\label{hWCLs:d}\\[2ex]
\Lambda_5 = x, &&
\DTM(x U_t) - \DHM\left( x_{+} U_x - U_{+}\right) = 0,\label{hWCLs:five}\\[2ex]
\Lambda_7 = t \dfrac{U_x + U_{\bar{x}}}{2} +x \dfrac{U_t + \check{U}_t}{2}, &&
\DTM\left(\dfrac{t+\hat{t}}{2}\dfrac{U_x+U_{\bar{x}}}{2}U_t + \dfrac{x}{2} U_t^2  + \dfrac{3x-x_+}{4}\hat{U}_{\bar{x}}{U}_{\bar{x}}
 \right) \label{hWCLs:six} \\[2ex]
&&
\hspace*{-1cm}
- \DHM\left(\dfrac{x+x_+}{2}U_x \dfrac{U_t+\check{U}_{t}}{2}  + \dfrac{3t-\hat{t}}{4} {\check{U}_t^+}\check{U}_t  + \dfrac{t}{2} U_x^2
 \right)  = 0, \nonumber
\end{eqnarray}
\end{subequations}
parallel to the conservation laws \eqref{CLsW} in the continuous case.

\begin{remark}
Note that the transformation generated by~$\sg{X}_7$ breaks the mesh orthogonality, and is not a symmetry of the difference equation \eqref{W:CLFORM} on the cross stencil. To show this, we consider the scheme~\eqref{W:CLFORM}
under the transformation corresponding to ~$\sg{X}_7$:
\begin{equation}\label{X7group_tr}
\def\arraystretch{1.5}
\begin{array}{c}
    t^* = t \cosh a + x \sinh a,
    \qquad
    \hat{t}^* = \hat{t} \cosh a + x \sinh a,
    \qquad
    \check{t}^* = \check{t} \cosh a + x \sinh a,
    \\
    x^* = x \cosh a + t \sinh a,
    \qquad
    x_+^* = x_+ \cosh a + t \sinh a,
    \qquad
    x_-^* = x_- \cosh a + t \sinh a,
    \\
    U^* = u(t^*, x^*),
    \quad
    U_+^* = u(t^*, x_+^*),
    \quad
    U_-^* = u(t^*, x_-^*),
    \quad
    \hat{U}^* = u(\hat{t}^*, x^*),
    \quad
    \check{U}^* = u(\check{t}^*, x^*).
\end{array}
\end{equation}
In \eqref{X7group_tr}, $a$ is the group parameter. Consider the points
\[
\def\arraystretch{1.5}
\begin{array}{l}
    A^* = (x \cosh a + t \sinh a, t \cosh a + x \sinh a),
    \\
    B^* = ((x + h) \cosh a + t \sinh a, t \cosh a + (x + h) \sinh a),
    \\
    C^* = (x \cosh a + (t + \tau) \sinh a, (t + \tau) \cosh a + x \sinh a)
\end{array}
\]
(Figure~\ref{fig:X7tr}). While the vectors ${AB}$ and ${AC}$ are orthogonal, the inner product
\[
({A^*B^*}, {A^*C^*}) =
(h \cosh a, h \sinh a) \cdot (\tau \sinh a, \tau \cosh a) = h \tau \sinh {2a} \neq 0,
\]
illustrating the breaking of the mesh orthogonality when $a\ne 0$. A sample transformation of a uniform orthogonal mesh by the transformation group \eqref{X7group_tr} corresponding to $\sg{X}_7$ is shown in Figure~\ref{fig:X7tr_mesh}.
\begin{figure}[ht]
\begin{center}
\includegraphics[width=0.5\linewidth]{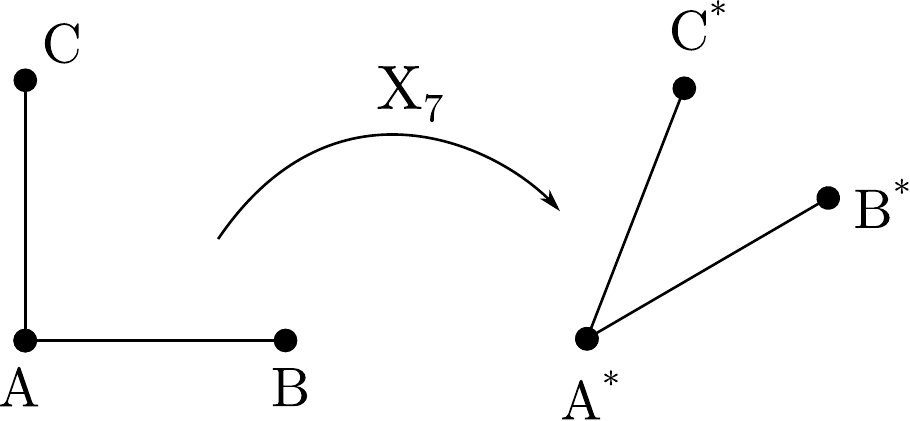}
\end{center}
\caption{Transformation of orthogonal vectors ${AB}$ and ${AC}$ under the action of the generator~$X_7$}
\label{fig:X7tr}
\end{figure}

\begin{figure}[ht]
\begin{center}
  \begin{minipage}[c]{0.4\linewidth}
    {\includegraphics[width=\linewidth]{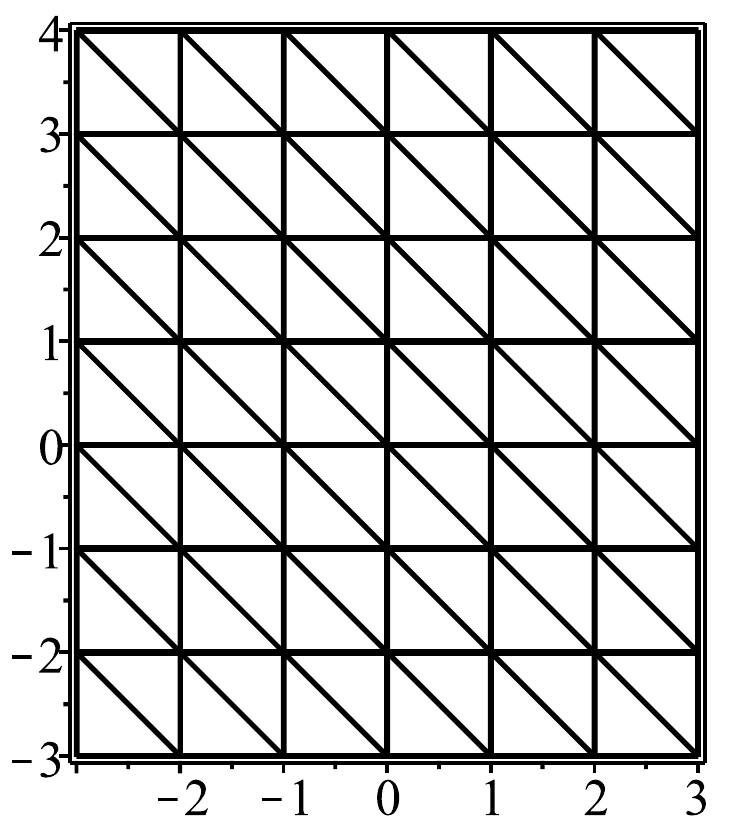}}
  \end{minipage}
  \hfil
  \begin{minipage}[c]{0.4\linewidth}
    \includegraphics[width=\linewidth]{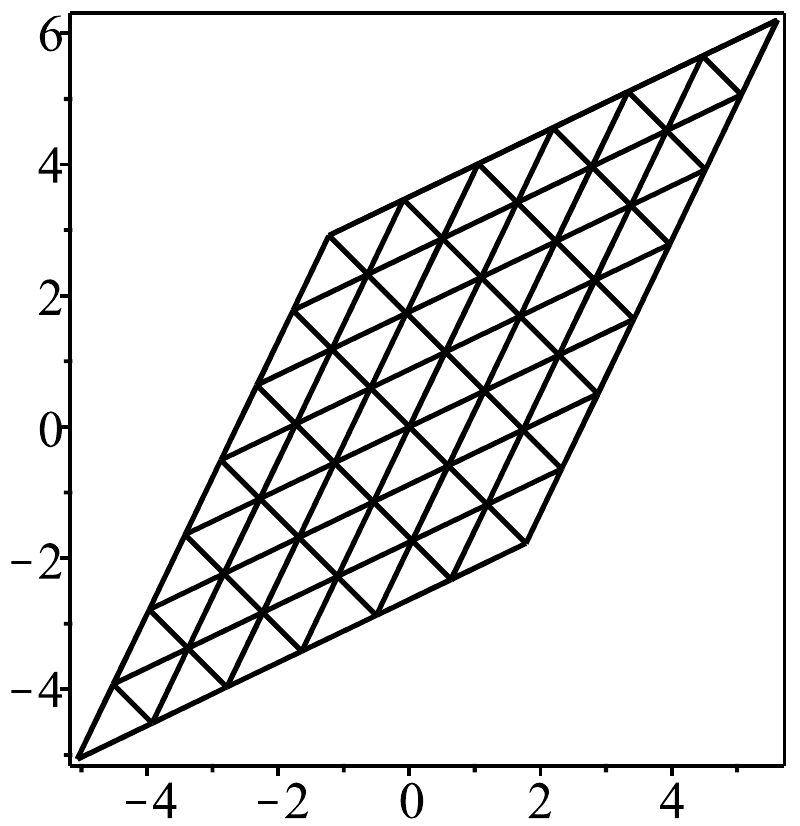}
  \end{minipage}
\end{center}
\caption{Transformation of a uniform orthogonal mesh~($h = \tau$)
under the action of the generator~$X_7$ corresponding to the group parameter~$a = \pi/6$. Left: the original mesh. Right: the transformed mesh.}
\label{fig:X7tr_mesh}
\end{figure}

Denote the angle between the transformed vectors ${A^*B^*}$ and ${A^*C^*}$ by  $\gamma$. Then
\[
    \cos\gamma = \tanh{2a},\quad \tan\gamma = \frac{1}{\sinh{2a}}.
\]
From Figure~\ref{fig:X7tr} one observes that $\hat{h}^* = {\tau^*}/{\tan \gamma} = \tau^* \sinh{2a}$. Expanding $\hat{U}^*$, $\check{U}^*$, ${U}^*_{\pm}$ in \eqref{X7group_tr} into series at some smooth solution~$u(t^*,x^*)$, one obtains
\[
\displaystyle
\def\arraystretch{1.5}
\begin{array}{c}
  \hat{U}^* = u + (u_t + u_x \sinh{2a})\tau^*
    + \frac{(\tau^*)^2}{2}(u_{tt} + u_{xx}\sinh^2{2a} + 2 u_{tx} \sinh{2a}) + O((\tau^*)^2),
  \\
  \check{U}^* = u - (u_t + u_x \sinh{2a})\tau^*
    + \frac{(\tau^*)^2}{2}(u_{tt} + u_{xx} \sinh^2{2a} + 2 u_{tx} \sinh{2a}) + O((\tau^*)^2),
  \\
  {U}^*_{\pm} = u \pm u_x h^* + \frac{(h^*)^2}{2} u_{xx} + O((h^*)^2).
\end{array}
\]
The transformed difference equation \eqref{W:CLFORM} on the cross stencil then reads
\[
  \frac{\hat{U}^* - 2 U^* + \check{U}^*}{(\tau^*)^2}
    -
    \frac{U^*_+ - 2 U^* + U^*_-}{(h^*)^2}
    =
    u_{tt} - (1 - \sinh^2{2a}) u_{xx}
    + 2 u_{tx} \sinh{2a} + O((h^*)^2 + (\tau^*)^2).
\]
The right side of the latter relation coincides with the linear wave equation (up to the error term $O((h^*)^2 + (\tau^*)^2)$) only in the case when $a=0$.
\end{remark}

\begin{remark}
For the linear wave equation \eqref{W}, the majority of the basic ``geometrical" symmetries \eqref{W:sym:geom} have corresponding conservation laws. For the finite-difference scheme \eqref{Wscheme} approximated on a cross-type five-point stencil on the uniform mesh, things are somewhat different. Indeed, \eqref{Wscheme} is a symmetry-invariant discretization with respect to all of the symmetries \eqref{W:sym:geom} and \eqref{eq:W:sym:alpha}, including the scaling $\sg{X}_6$, except for the Lorentz symmetry $\sg{X}_7$ that breaks the mesh orthogonality. Yet in the set of corresponding discrete conservation laws with multipliers defined on the cross stencil, we find an analog of the conservation law corresponding to the symmetry $\sg{X}_7$ (multiplier $\Lambda_7$) but not the scaling symmetry $\sg{X}_6$!
\end{remark}

%


\section{Example 2: the Nonlinear Wave Equation}\label{sec:nonlinW}


As a nonlinear example, consider the wave equation \eqref{eq:anz:1fib:hyperel:gam0:PDE} describing general transverse shear waves in a hyperelastic fiber-reinforced material. Using a scaling transformation
\[
x={\hat{x}},\qquad t=\alpha^{-1/2}{\hat{t}},\qquad G(x,t)=\Big(\dfrac{\alpha}{3\beta}\Big)^{1/2}\hat{u}({\hat{x}},{\hat{t}})
\]
and dropping the hats, a simpler PDE form is obtained:
\begin{equation} \label{Wn}
    W_{\text{NL}} = u_{tt} - (1 + u_{x}^2) u_{xx} = 0,
\end{equation}
involving no parameters. It is straightforward to show that the PDE \eqref{Wn} admits five point symmetries with infinitesimal generators
\begin{equation} \label{WnSyms}
\sg{Y}_1 = \frac{\partial}{\partial u}, \quad
\sg{Y}_2 = \frac{\partial}{\partial x}, \quad
\sg{Y}_3 = \frac{\partial}{\partial t}, \quad
\sg{Y}_4 = t \, \frac{\partial}{\partial u}, \quad
\sg{Y}_5 = x \, \frac{\partial}{\partial x} + t \, \frac{\partial}{\partial t} + u \, \frac{\partial}{\partial u}.
\end{equation}
%
%
Moreover, \eqref{Wn} is the Euler-Lagrange equation for the Lagrangian
\begin{equation}\label{WnL}
  L_N = \dfrac{1}{2}(u_x^2 - u_t^2) + \frac{1}{12}u_x^4.
\end{equation}
Since all symmetries but $\sg{Y}_5$ preserve the corresponding mechanical action, by the Noether's first theorem, the evolutionary forms of the generators $\sg{Y}_1$,\ldots,$\sg{Y}_4$ yield conservation law multipliers. These multipliers and the corresponding local conservation laws are given by
\begin{subequations}\label{CLsWn}
\begin{eqnarray}
  M_1 = 1, & D_t\left( u_t \right) - D_x\left( u_x + \dfrac{u_x^3}{3} \right) = 0, \label{CLsWn:sub1} \\
  M_2 = u_x, & D_t\left( u_t u_x \right) - D_x\left( \dfrac{u_t^2 + u_x^2}{2} + \dfrac{u_x^4}{4}\right) = 0, \label{CLsWn:sub2}\\
  M_3 = u_t, & D_t\left( \dfrac{u_t^2 + u_x^2}{2} + \dfrac{u_x^4}{12}\right) - D_x\left( u_t u_x \left( 1 + \dfrac{u_x^2}{3}\right) \right) = 0, \label{CLsWn:sub3}\\
  M_4 = t,  & D_t\left(t u_t - u \right) - D_x\left( t \left( u_x + \dfrac{u_x^3}{3} \right) \right) = 0. \label{CLsWn:sub4}
\end{eqnarray}
\end{subequations}
In particular, the first and the third one correspond to the local conservation of momentum and energy, and the fourth one describes the motion of the center of mass (cf. \eqref{cmass}, \cite{cheviakov2016one}).

\subsection{The cross-type stencil for the nonlinear wave equation}

Let us now consider difference analogs of  the nonlinear wave equation \eqref{Wn}.  First, consider the five-point cross-type stencil on the orthogonal uniform mesh (Fig.~\ref{fig:5pt-stencil}). Similarly to the linear case, it is straightforward to find a finite-difference scheme possessing two difference conservation laws, corresponding to the symmetries $\sg{Y}_1$ and $\sg{Y}_4$ in \eqref{WnSyms}, and respectively to the continuum conservation laws \eqref{CLsWn:sub1} and \eqref{CLsWn:sub4}.
There are infinitely many such schemes (because of infinitely many divergence approximations
of expression~$D_x(u_x^3)$ that admit multipliers~$1$~and~$t$, e.~g.
$\DHM(\theta U_x^3 + (1 - \theta)\hat{U}_x^3)$, where $0 \leqslant \theta \leqslant 1$).
An example is given by
\begin{equation} \label{Scheme2CLs}
\begin{cases}
  U_{t\check{t}} - U_{x\bar{x}} - \dfrac{U_x^3 - U_{\bar{x}}^3}{3 h} = 0, \\
  h^+ = h^- = h = \const, \\
  \hat{\tau} = \check{\tau} = \tau = \const
\end{cases}
\end{equation}
with the following local difference conservation laws and corresponding multipliers:
\begin{eqnarray}
&\Lambda_1 = 1, &  \DTM(U_t) - \DHM\left(U_x + \frac{U_x^3}{3}\right) = 0, \\
&\Lambda_4 = t, &  \DTM(t U_t - U) - \DHM\left(t U_x + t \frac{U_x^3}{3}\right) = 0.
\end{eqnarray}

However, one can show that for the cross-type stencil, there are no polynomial schemes with more than two conservation laws; those conservation laws will always correspond to the first and the fourth continuum conservation laws \eqref{CLsWn:sub1} and \eqref{CLsWn:sub4}, as in the scheme \eqref{Scheme2CLs}. For the cross-stencil, there also exist schemes with only one difference conservation law, corresponding to the continuum conservation law \eqref{CLsWn:sub2}. (Below we give such an example for the nine-point stencil.)

\subsection{Construction of a scheme with three conservation laws}\label{sec:NLW:constr3}

\begin{figure}[h]
\centering
\includegraphics[scale=0.75]{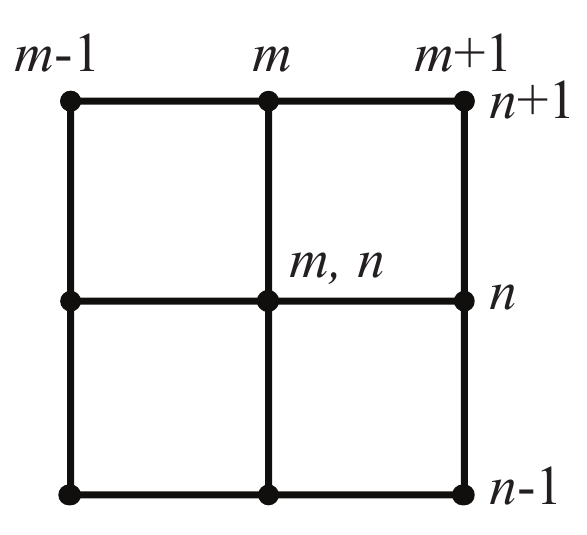}
\caption{The nine-point stencil.}
\label{fig:9pt-stencil}
\end{figure}

In order to find a difference analog of the PDE \eqref{Wn} possessing more than two local difference conservation laws, one must extend the stencil. Let us seek such schemes on the nine-point stencil  on the uniform orthogonal  mesh (Fig.~\ref{fig:9pt-stencil}).
The finite-difference scheme is then written using the mesh variables
\beq\label{template}
\barr
(\mathbf{t}, \mathbf{x}, \mathbf{U})
\equiv \left(
        \check{t}, t, \hat{t},
        x_-, x, x_+,
        \check{U}_-, \check{U}, \check{U}_+,
        U_-, U, U_+,
        \hat{U}_-,  \hat{U}, \hat{U}_+
    \right) {} \\[2ex]
  \qquad = \left(
        t_{n-1}, t_n, t_{n+1},
        x_{m-1}, x_{m}, x_{m+1},\right.\\[2ex]
  \qquad \quad      \left. U^{n-1}_{m-1}, U^{n-1}_{m}, U^{n-1}_{m+1},
        U^{n}_{m-1}, U^{n}_{m}, U^{n}_{m+1},
        U^{n+1}_{m-1}, U^{n+1}_{m}, U^{n+1}_{m+1}
  \right).
  \earr
\eeq

Since the PDE \eqref{Wn} is a differential polynomial in terms of $u_x, u_{xx}, u_{tt}$, we will seek a finite-difference scheme representable by some polynomial on the stencil \eqref{template}. We demand the symmetry invariance of the scheme; the symmetry $\sg{Y}_1$ \eqref{WnSyms} corresponds to the integrating factor $\Lambda_1 = 1$ for the difference equation, therefore the latter must have a divergence form. So we seek the scheme as a difference divergence expression
\begin{equation} \label{Wgeneral}
F(\mathbf{t}, \mathbf{x}, \mathbf{U}) \equiv \WNL = \DTP(\Theta) + \DHP(\Phi) = 0,
\end{equation}
where $\Theta$ and  $\Phi$ are difference polynomials approximating the density and the flux of the local continuous conservation law \eqref{CLsWn:sub1}. From the form of \eqref{CLsWn:sub1} it is clear that $\Theta$ is a linear expression, and $\Phi$ is some cubic polynomial.

Let us seek the conserved density $\Theta$ as a generic linear expression depending on six points of the stencil, i.e., depending $U$,$U_+$,$U_-$,$\check{U}$,$\check{U}_+$,$\check{U}_-$, so that the operator $\DTP$ does not take us out of the nine-point stencil:
\begin{equation}\label{psiT}
  \Theta = a_1 U + a_2 U_+ + a_3 U_- + a_4 \check{U} + a_5 \check{U}_+ + a_6 \check{U}_-;
\end{equation}
here $\mathbf{a} = (a_1, \dots, a_6)$ are arbitrary constant coefficients. Similarly, we will seek $\Phi$ in the form of a cubic polynomial depending on the mesh quantities $U$,$\hat{U}$, $\check{U}$, $U_-,\hat{U}_-$, $\check{U}_-$:
\beq \label{psiX}
\barr
  \Phi = A_1 U^3 + \dots + A_6 (\check{U}_-)^3 \\
  \quad + B_1^1 U^2 \hat{U} + B_1^2 U \hat{U}^2 + \dots + B_{15}^1 (\hat{U}_-)^2 \check{U}_- + B_{15}^2 \hat{U}_- (\check{U}_-)^2 \\
  \quad + C_1 U \hat{U} \check{U} + \dots + C_{20} U_- \hat{U}_- \check{U}_- \\
  \quad + E_1 U^2 + \dots + E_6 (\check{U}_-)^2 \\
  \quad + F_1 U \hat{U} + \dots + F_{15} \hat{U}_- \check{U}_-\\
  \quad + G_1 U + \dots + G_6 \check{U}_-,\\
\earr
\eeq
where
$\mathbf{A} = (A_1, \dots, A_6)$,
$\mathbf{B} = (B_1^1, \dots,  B_{15}^1$, $B_1^2, \dots,  B_{15}^2)$,
$\mathbf{C} = (C_1, \dots, C_{20})$,
$\mathbf{E} = (E_1, \dots, E_6)$,
$\mathbf{F} = (F_1, \dots, F_{15})$,
$\mathbf{G} = (G_1, \dots, G_6)$
are arbitrary real vector coefficients. The total number of these coefficients is 83. For the spatial flux form \eqref{psiX} we again used the six points such that $\DHP(\Phi)$ stays within the nine-point stencil. Let us also seek conservation law multipliers in the form of linear functoins
\begin{equation}\label{mult}
  \Lambda_z = z_1 U + z_2 U_+ + z_3 U_- + z_4 \hat{U} + z_5 \hat{U}_+
  + z_6 \hat{U}_- + z_7 \check{U} + z_8 \check{U}_+ + z_9 \check{U}_-,
\end{equation}
where $\mathbf{z} = (z_1, \dots, z_9)$ are arbitrary constant coefficients.

In summary, we seek such constants $\mathbf{a}$, $\mathbf{A}$, \ldots, $\mathbf{G}$, $\mathbf{z}$ that the conditions
\eqref{detSys0}--\eqref{detSys0Eu} are satisfied, and moreover, \eqref{Wgeneral} approximates the nonlinear PDE \eqref{Wn}.

Taking into account that for the operators \eqref{WnSyms}, all conditions  \eqref{latticeConstraints1} are satisfied, and that the equations of the divergence form \eqref{Wgeneral} (with density and flux not explicitly depending on $x,t$) are always invariant with respect to space and time shifts $\sg{Y}_2$, $\sg{Y}_3$ in \eqref{WnSyms}, we write the equations \eqref{detSys0}--\eqref{detSys0Eu} in the form
\begin{subequations}\label{detSysWnAll}
\begin{eqnarray}\label{detSysWn}
\WNL = \DTP(\Theta) + \DHP(\Phi) = 0, \label{WNL:scheme}\\
h_+ = h_- = h, \label{WNL:scheme2}\\
\check{\tau} = \hat{\tau} = \tau,\label{WNL:scheme3} \\
(\underset{h}{\text{pr}}~\sg{Y}_1)\, \WNL
    \equiv \left(\underset{h}{\text{pr}} \, \frac{\partial}{\partial u}\right)\,  \WNL|_{[\Delta_h]}   = 0, \label{detSysWnX} \\
(\underset{h}{\text{pr}}~\sg{Y}_4)\, \WNL
    = \left(\underset{h}{\text{pr}} \, t \frac{\partial}{\partial u}\right)\,  \WNL|_{[\Delta_h]}   = 0, \label{detSysWnXt} \\
\mathcal{E}_U ( \Lambda_z \, \WNL ) \equiv 0, \label{detSysWnEu}
\end{eqnarray}
\end{subequations}
where \eqref{detSysWnX}, \eqref{detSysWnXt} hold on solutions of $\Delta_h$: \eqref{WNL:scheme}--\eqref{WNL:scheme3}, and \eqref{detSysWnEu} holds identically for any mesh quantity. After splitting according to the powers of mesh quantities, the equations \eqref{detSysWnX}--\eqref{detSysWnEu} lead to a bulk system containing several hundred bilinear equations for the unknown coefficients $\mathbf{a}$, $\mathbf{A}$\ldots,$\mathbf{G}$, $\mathbf{z}$. This system is consistent, and has multiple solutions, each yielding a multiplier-difference equation pair
\begin{equation} \label{pairs}
  \left(\Lambda_z^k, \, \WNL^k \right), \qquad k=1,2, \dots.
\end{equation}
Some of such solutions tend to zero in the continuum limit. Here we need to find multipliers $\Lambda_z$ \eqref{mult} that in the continuum limit tend to the partial derivatives $u_t$, $u_x$ or linear combinations thereof. These multipliers would correspond to  the conservation laws \eqref{CLsWn:sub2}, \eqref{CLsWn:sub3}.


We start from the conservation law \eqref{CLsWn:sub2}. Considering different pairs  \eqref{pairs}, where the multipliers $\Lambda_z^k$ become  $u_x$ in the continuum limit, one notices that the corresponding difference equations $\WNL^k$ approximate only \emph{linear} equations. The coefficients $\mathbf{A}$, $\mathbf{B}$, $\mathbf{C}$, $\mathbf{E}$, $\mathbf{F}$ are responsible for the nonlinear terms in the difference equation $\WNL$ \eqref{detSysWn}. It follows from \eqref{detSysWnAll} that when any of these constants are nonzero, the coefficients  $z_2$, $z_3$, $z_5$, $z_8$, $z_9$ in the multipliers  \eqref{mult} will vanish. In this case,  it is not possible that the multiplier expression \eqref{mult} will yield $u_x$ in the continuum limit. Based on that, we conclude that there is \emph{no symmetry-invariant polynomial finite-difference scheme \eqref{detSysWnAll} on the nine-point stencil \eqref{template} that approximates the nonlinear wave equation  \eqref{Wn} and possesses a difference analog of the local conservation law  \ref{CLsWn:sub2}).} [The above argument is based on direct verification of various solutions of a large system of equations; perhaps there is a more elegant proof that would be based on some qualitative properties of the PDE and the stencil.]

\medskip

We now consider the local conservation law \eqref{CLsWn:sub3} of the nonlinear PDE $W_{\text{NL}}$ \eqref{Wn}. Here one can find a multitude of multiplier-difference equation pairs \eqref{pairs} where the difference equation $\WNL^k$ possesses an analog of \eqref{CLsWn:sub3}. For example, one can choose the multiplier to be
\begin{equation}
  \Lambda_3 = \dfrac{\hat{U} - \check{U}}{2 \tau} \equiv \frac{U_t + \check{U}_t}{2},
\end{equation}
which is the same as the corresponding multiplier \eqref{eq:LinW:Lam23} for the linear wave equation. This particular form is attained when $z_4 = 1 / (2 \tau)$, $z_7 = -z_4$, and $z_i = 0$ for $i \neq 4, \, 7$. At these values of constants, the equation \eqref{detSysWnEu} turns out to be linear, and has a solution
\beq \label{WnRaw}
\barr
\WNL = \dfrac{1}{\tau} \, \Big[  a_{{3}}{\hat{U}_-} -a_{{5}}{\check{U}_-} + \left( a_{{5}} -a_{{3}} \right) {U_-}+{\check{U}_+}a_{{3}} \\[2ex]
\qquad     -a_{{5}}{\hat{U}_+}+{U_+} \left( a_{{5}} -a_{{3}}\right) + \left( a_{{4}}+a_{{6}}+a_{{3}} \right){\check{U}} \\[2ex]
    \qquad + \left( a_{{4}}+a_{{6}}+a_{{3}} \right) {\hat{U}} - 2 \left(a_{{4}}+a_{{6}}+a_{{3}} \right) U
\Big] + \dfrac{1}{h} \, \Big[ \dots \Big],
\earr
\eeq
where the final brackets contain a bulk cubic polynomial expression. In order to compute the remaining unknown coefficients, we use the method of undetermined coefficients \cite{godunov1973raznostnye}. It consists in replacing the approximate mesh quantities with their exact counterparts, such as $U=U_{m}^{n}\to u(t_n,x_m)$, etc.,  expanding the solution in power series in terms of $\tau$ and $h$ about the center of the stencil, and using the resulting expression find out the remaining coefficients. We obtain
\beq \label{series}
\barr
\WNL =
6\, {h}^{3} \left( 3\,B_4^2 +B_{13}^1 - B_{15}^2 + 2\,C_{{20}} \right) u_x^2 u_{xx} + 2\,{h}^{2} \left( E_{{13}}+2\,F_{{5}}+3\,F_{{6}} \right) u_x u_{xx} \\[2ex]
\quad -h \, (G_3+G_4+G_5) u_{xx} + \left( {\dfrac { \left( G_{{3}}+G_{{6}} \right) {\tau}^{2}}{h}}+ \left( 2\,a_{{3}}+a_{{4}}-a_{{5}}+a_{{6}} \right) \tau \right) \, u_{tt} \\[2ex]
\quad + 2 \, \left(
    24\,{\tau}^{3} \left( B_{{4,2}}+B_{{13,1}} \right)  u_t^2
    - 4\,{\tau}^{2} \left( E_{{13}}-F_{{5}} \right) u_t
     - \left(G_{{4}} - G_{{5}} \right) \tau
     - \left( a_{{3}} + a_{{5}} \right) h
\right) u_{tx} \\[2ex]
\quad + \dfrac{1}{12} \,\left(
    {\dfrac { \left( G_{{3}}+G_{{6}} \right) {\tau}^{4}}{h}}
    + \left( 2\,a_{{3}}+a_{{4}}-a_{{5}}+a_{{6}} \right) {\tau}^{3}
\right) u_{tttt} \\[2ex]
\quad - \dfrac{1}{12} \,{h}^{3} \left( G_{{3}}+G_{{4}}+G_{{5}} \right) u_{xxxx} \\[2ex]
\quad - \dfrac{1}{3} \, \left( (G_4 - G_5) \tau^3 + (a_3 + a_5) h \tau^2 \right) u_{xttt} \\[2ex]
\quad - \dfrac{1}{2} \, \left( (G_4 + G_5) h \tau^2 + (a_5 - a_3) h^2 \tau \right) u_{ttxx} \\[2ex]
\quad - \dfrac{1}{3} \left( (G_4 - G_5) h^2 \tau + (a_3 + a_5) \right) u_{xxxt}  \\[2ex]
\quad + \mathcal{O}\left((\tau + h)^4\right).
\earr
\eeq
This leads to the constant choices
\begin{eqnarray}
  B_4^2 = \frac{1}{2} \, B_{15}^2 - C_{20} + \frac{1}{12 h^3}, \quad
  E_{13} = F_5 = -F_6, \nonumber \\
  G_4 = G_5 = a_3 = a_5 = 0, \quad
  G_3 = -G_6, \\
  G_6 = 1/h, \quad
  a_4 = -a_6 - 1/\tau. \nonumber
\end{eqnarray}
The coefficients $B^1_{13}, C_{20}, F_6$ influence only the higher-order terms in $h$ and $\tau$, and therefore can be chosen arbitrary, for example, set to zero.

After the substitution of the above into \eqref{WnRaw}, the finite-difference PDE approximation takes the form
\[
    \WNL = U_{t\check{t}} - U_{x\bar{x}} - \frac{1}{6} \DHM\left( U_x^2 \, (\hat{U}_x + \check{U}_x)\right) = 0,
\]
and the scheme is given by
\begin{equation} \label{Scheme3CLs}
  \begin{cases}
    \WNL = U_{t\check{t}} - U_{x\bar{x}} - \dfrac{1}{6} \DHM\left( U_x^2 \, (\hat{U}_x + \check{U}_x)\right) = 0, \\
    h_+ = h_- = h = \text{const}, \\
    \hat{\tau} = \check{\tau} = \tau = \text{const}.
  \end{cases}
\end{equation}
The finite-difference scheme \eqref{Scheme3CLs} is an implicit scheme providing a second-order approximation of the nonlinear PDE \eqref{Wn} in  $h$ and $\tau$; this scheme is invariant with respect to discrete symmetries corresponding to all five point symmetry generators \eqref{WnSyms} of the continuum model PDE \eqref{Wn}. The approximation \eqref{Scheme3CLs} admits three discrete conservation laws with the above-discussed multipliers, from which discrete densities and fluxes are determined by analogy with the linear case \eqref{Wscheme}, \eqref{hWCLs}:
\begin{subequations}\label{eq:nonlW:3CLs:forms}
\begin{eqnarray}
&\Lambda_1 = 1, &
\DTM(U_t) - \DHM\left( U_x \left(1 + U_x \, \frac{\hat{U}_x + \check{U}_x}{6} \right) \right) = 0, \\
&\Lambda_3 = \dfrac{U_t + \check{U}_t}{2}, &
\DTM\left(
    \dfrac{U_x \hat{U}_x + U_t^2}{2} + \frac{U_x^2 \hat{U}_x^2}{12}
\right) \nonumber \\
 && \qquad - \DHM\left(
    U_x \, \frac{U_t^+ + \check{U}_t^+}{2} \, \left(
        1 + U_x \, \frac{\hat{U}_x + \check{U}_x}{6}
    \right)
\right) = 0, \\
&\Lambda_4 = t, &
\DTM(t U_t - U) - \DHM\left( t\, U_x \left(1 + U_x \, \frac{\hat{U}_x + \check{U}_x}{6} \right) \right) = 0.
\end{eqnarray}
\end{subequations}
Importantly, while implicit, the scheme \eqref{Scheme3CLs} is linear in the mesh values $\hat{U}_-$, $\hat{U}$, $\hat{U}_+$ of the unknown (upper) time layer in the nine-point stencil (Figure \ref{fig:9pt-stencil}). It follows that in the time-stepping, one can use not only linear methods like Gaussian elimination, but also the numerically more efficient tridiagonal solver.

\subsection{Schemes admitting the analog of the conservation law \eqref{CLsWn:sub2}}\label{sec:nonlW:missing}

We have determined above that there is no invariant polynomial scheme of the form \eqref{Wgeneral}, approximating the nonlinear PDE \eqref{Wn}  on the nine-point stencil (Fig.~\ref{fig:9pt-stencil}) and possessing a difference analog of the conservation law  \eqref{CLsWn:sub2}.

We now weaken the requirements, and present an example of a finite-difference scheme with one conservation law corresponding to \eqref{CLsWn:sub2}. Let us seek the scheme in a non-divergece form, i.e., instead of the expression  \eqref{Wgeneral}, attempt to obtain  some other polynomial difference expression   $\WNL^*$ which would satisfy the condition \eqref{detSysWnEu}
\begin{equation}
  \mathcal{E}_U ( \Lambda_z^* \, \WNL^* ) \equiv 0,
\end{equation}
where $\Lambda_z^*$ is again sought in the form  \eqref{mult}, and in the continuum limit, would correspond to the multiplier $M_2 = u_x$ of the ``missing'' conservation law
\eqref{CLsWn:sub2}. Repeating basically the same procedure used to obtain the scheme \eqref{Scheme3CLs}, we obtain a non-divergent scheme
\begin{equation} \label{Scheme1CL}
  \begin{cases}
    U_{t\check{t}} - U_{x\bar{x}} - \dfrac{U_x^2 + U_{\bar{x}}^2}{2} \, U_{x\bar{x}} = 0, \\
    h_+ = h_- = h = \text{const}, \\
    \hat{\tau} = \check{\tau} = \tau = \text{const}.
  \end{cases}
\end{equation}
This scheme is invariant with respect to Lie point symmetries given by \eqref{WnSyms} and admits a conservation law
\begin{equation}
    \Lambda^*_2 = \frac{U_x + U_{\bar{x}}}{2}, \qquad
    \DTM\left(U_t \frac{\hat{U}_x + \hat{U}_{\bar{x}}}{2}\right) - \DHM\left( \frac{U_t U_t^+ + U_x^2}{2} + \frac{U_x^4}{4} \right) = 0.
\end{equation}
analogous to \eqref{CLsWn:sub2}. The scheme \eqref{Scheme1CL} is explicit, defined on the cross-type stencil over the uniform orthogonal  mesh (Fig.~\ref{fig:5pt-stencil}).

\begin{remark}
In a similar way, one can obtain other schemes with a conservation law corresponding to \eqref{CLsWn:sub2}. For example, the integrating factor
\begin{equation}
\Lambda= \frac{\hat{U}_x + \hat{U}_{\bar{x}} + \check{U}_x + \check{U}_{\bar{x}}}{4},
\end{equation}
yields an implicit scheme defined on a nine-point stencil (Fig.~\ref{fig:9pt-stencil}). [This scheme, however, has no obvious advantages over \eqref{Scheme1CL}.]
\end{remark}


\section{Discussion}  \label{conclusion}

In addition to the approximation order, the notion of ``quality" of approximation provided by a finite-difference numerical method may also include certain analytical and geometrical properties reflecting those of the given PDE model. In the current work, we considered the self-adjoint linear and nonlinear (1+1)-dimensional PDEs \eqref{W} and \eqref{Wn}, their point symmetries, and local conservation laws, and the question of systematic construction of finite-difference approximations preserving that structure. In particular, on a general five- or nine-point stencil in a uniform orthogonal mesh, one can construct infinitely many discretizations approximating a given second-order PDE; using discrete symmetry prolongation and the discrete analog of the direct conservation law construction method (Section \ref{sec:2}), we used the requirement of symmetry invariance and existence of corresponding conservation law multipliers to derive discretizations that preserve additional qualitative features of the original differential model.
%

The linear homogeneous constant-coefficient wave equation \eqref{W} admits an infinite number of Lie point symmetries and conservation laws, including those corresponding to its linearity (Section \ref{sec:linW}). In particular, point symmetry components satisfy \eqref{W:symGen:eta}, \eqref{W:symGen:Xis:Eq}, and first-order conservation law multipliers are arbitrary solutions of the linear system \eqref{eq:linW:CL:Mult:DetEq}. Restricting to the seven basic geometrical symmetries \eqref{W:sym:geom}, and the corresponding six conservation laws \eqref{CLsW}, we pose a question of constructing a simple finite-difference scheme that admits similar symmetries and conservation laws. We show that the simplest central difference five-point scheme \eqref{Wscheme} on a symmetric stencil within a unform mesh (Figure \ref{fig:5pt-stencil}) admits analogs of all geometrical symmetries \eqref{W:sym:geom} except for the Lorentz boost $\sg{X}_6$ (which does not preserve the mesh properties), including the infinite set of linearity symmetries. In terms of conservation laws, the central difference scheme \eqref{Wscheme} admits the analogs \eqref{hWCLs} of five out of six conservation laws \eqref{CLsW}. Interestingly, for the missing symmetry $\sg{X}_6$, a discrete conservation law with an analogous multiplier $\Lambda_6$ arises; on the contrary, for the scaling symmetry $\sg{X}_5$ inherited by the discretization, the discrete conservation law analogous to \eqref{CLsW:sub5} does not arise.

In Section \ref{sec:nonlinW}, discretizations of the nonlinear wave equation \eqref{Wn} were considered. This PDE is variational with a Lagrangian \eqref{WnL}, admits five point symmetries \eqref{Scheme2CLs}, and four corresponding local conservation laws \eqref{CLsWn}. We then considered discretizations of the nonlinear wave equation \eqref{Wn} in light of their invariance and conservation properties. We showed that for explicit invariant schemes on the cross-type stencil in a uniform mesh, there are no polynomial schemes with more than two conservation laws. For example, a simple scheme \eqref{Scheme2CLs} admits discrete analogs of first and fourth conservation laws \eqref{CLsWn}. In Section \ref{sec:NLW:constr3}, imposing the symmetry invariance and divergence form requirements, we construct a scheme \eqref{Scheme3CLs} on the nine-point stencil, admitting three discrete conservation laws -- analogs of first, third and fourth conservation laws \eqref{CLsWn} of the PDE model \eqref{Wn}. The scheme \eqref{Scheme3CLs} is an implicit second-order scheme, linear in the unknown upper-layer values of $U$, so that the tridiagonal solver can be used for time stepping.

It was shown that there is no invariant polynomial scheme of the conserved form \eqref{Wgeneral}, approximating the nonlinear PDE \eqref{Wn}  on the nine-point stencil \eqref{template} (Fig.~\ref{fig:9pt-stencil}) and possessing a difference analog of the conservation law  \eqref{CLsWn:sub2}. Many schemes admitting the analog of the missing conservation law \eqref{CLsWn:sub2} exist, but they would then fail to preserve other conservation laws (Section \ref{sec:nonlW:missing}).

In summary, for the nonlinear wave equation \eqref{Wn}, on a uniform orthogonal mesh \eqref{regLattice}, there are finite-difference invariant with respect to the symmetries \eqref{WnSyms} and belonging to three classes:
\begin{enumerate}
  \item Schemes with a single discrete conservation law corresponding to \eqref{CLsWn:sub2}. An example is given by an explicit scheme \eqref{Scheme1CL}.
  \item Discretizations with two discrete conservation laws corresponding to \eqref{CLsWn:sub1}, \eqref{CLsWn:sub4}. An example is provided by an explicit scheme \eqref{Scheme2CLs} on the five-point cross-type stencil.
  \item Schemes with three discrete conservation laws corresponding to \eqref{CLsWn:sub1}, \eqref{CLsWn:sub3}, and \eqref{CLsWn:sub4}, with an example is given by \eqref{Scheme3CLs}.
\end{enumerate}

We also note that using the direct method to compute discrete conservation law multipliers for finite-difference approximations, as done in Sections \ref{sec:linW} and \ref{sec:nonlinW}, one gets infinitely many multipliers $\Lambda$ that tend to zero as $\tau, h\to 0$. These multipliers yield discrete conservation laws that correspond to trivial conservation laws in the continuous PDE model. Such multipliers were not listed in the computations.

The approach presented in the current contribution can be generalized to multiple spatial dimensions, as well as to meshes that are non-uniform and/or non-orthogonal. (For the invariance criterion for such meshes and various examples, including situations where meshes explicitly depend on the solution, see, e.g., Ref.~\cite{[D-book]}.) We also note that the applicability of the method is not limited to polynomial finite-difference schemes. In particular, for schemes that can be represented in terms of rational functions, the problem can be reduced to the consideration of polynomials. For schemes that are not reducible to rational functions, the situation may be more complicated.

In future work, we intend to compare the performance of the numerical schemes, including \eqref{Scheme1CL}, \eqref{Scheme2CLs}, and the ``most conservative" discretization \eqref{Scheme3CLs} of the nonlinear wave equation \eqref{Wn}, in particular, study their numerical stability, error behaviour, and the relative ability to preserve discrete analogs of conservation laws \eqref{CLsWn} of the PDE \eqref{Wn}. From the theoretical point of view, it is of interest to work towards formulating a general method to systematically seek discretizations of PDEs and PDE systems preserving multiple discrete conservation laws, simultaneously with seeking the discrete forms of conservation laws themselves. Such an algorithm can be possibly also based on the direct (multiplier) approach which, importantly, always yields determining equations linear in the unknown conservation law multipliers.

\section*{Acknowledgements}

A.C. is grateful to NSERC of Canada for research support through the Discovery grant program.
E.K. is grateful to S.~V. Melesko for valuable discussions and sincerely
appreciates the hospitality of the Suranaree University of Technology.

\section*{Data Availability Statement}

Data sharing is not applicable to this article as no new data were created or analyzed in this study.

{\footnotesize
\bibliographystyle{ieeetr}
\bibliography{wavebib26}
}

\end{document}